\documentclass[aps,prb,twocolumn,10pt,showpacs,superscriptaddress]{revtex4-2}

\usepackage[utf8]{inputenc}
\usepackage[T1]{fontenc}

\usepackage{amsmath}
\usepackage{amssymb}
\usepackage{physics}
\usepackage{graphicx}
\usepackage{float}
\usepackage{booktabs}
\usepackage{siunitx}
\usepackage{xcolor}
\usepackage{orcidlink}
\usepackage{hyperref}

\hypersetup{
	colorlinks=true,
	linkcolor=blue,
	citecolor=blue,
	filecolor=blue,
	urlcolor=blue,
	breaklinks=true
}

\begin{document}
	
	\title{Screw-dislocation-engineered quantum dot: geometry-tunable nonlinear optics, orbital qubit addressability, and torsion metrology}
	
	\author{Edilberto O. Silva\,\orcidlink{0000-0002-0297-5747}}
	\email{edilberto.silva@ufma.br}
	\affiliation{Programa de P\'os-Gradua\c c\~ao em F\'{\i}sica \& Coordena\c c\~ao do Curso de F\'{\i}sica -- Bacharelado, Universidade Federal do Maranh\~{a}o, 65085-580 S\~{a}o Lu\'{\i}s, Maranh\~{a}o, Brazil}
	
	\date{\today}
	
	\begin{abstract}
		We study a single electron confined in a uniform-torsion medium, a continuum model of a screw dislocation density, in a perpendicular magnetic field, and in the presence of an Aharonov-Bohm flux. Torsion alone produces radial confinement without any \textit{ad hoc} potential, while the Aharonov–Bohm phase breaks the usual $m\leftrightarrow -m$ symmetry. From the exact spectrum and wave functions, we find: (i) a torsion-controlled optical transition whose energy blue-shifts from $\sim6.8$ to $\sim15.5$ meV and whose saturation intensity varies by an order of magnitude, enabling geometry-programmable optical switching; (ii) an Aharonov-Bohm-tunable ``angular pseudospin'' formed by the $m=\pm1$ states, with flux-controlled level splitting and asymmetric oscillator strengths that allow selective optical addressability; and (iii) an approximately linear torsion dependence of the transition energy that enables nanoscale torsion metrology with an estimated resolution of $\sim 10^{5}~\mathrm{m}^{-1}$. In this context, ``torsion'' refers to the experimentally relevant continuum limit of a uniform density of parallel screw dislocations, i.e., a crystal with finite torsion but vanishing curvature, which can, in practice, be engineered and probed in twisted nanowires and strained semiconductor heterostructures. We also show how torsion provides \textit{in situ} control of emitter–cavity detuning and light–matter coupling in cavity QED, in direct analogy with strain tuning of semiconductor quantum dots in nanocavities, but here arising from a purely geometric/topological parameter.
	\end{abstract}

	\maketitle
	
	\section{Introduction}
	\label{sec:introduction}
	
	Modern quantum technologies rely on actively tuning localized quantum states: shifting their transition energies, modifying their dipole strengths, selectively addressing one branch of a near-degenerate manifold, or reading out physical parameters (strain, fields, topology) from optical spectra \cite{Degen2017RMP,ChemlaShah2001Nature}. The established tuning knobs are well known: electrostatic gating, elastic strain, magnetic fields, and engineered confinement potentials. These approaches underpin solid-state quantum emitters, nonlinear nanophotonics, and defect-based quantum sensing. What remains largely unexplored is whether the \emph{geometry and topology of the host medium itself} can be promoted to programmable, device-level resources.
	
	Here, we show that uniform torsion, a continuum description of a parallel distribution of screw dislocations modeled as a space with finite torsion and vanishing curvature \cite{Katanaev1992AnnPhys,Furtado1999EPL}, is precisely such a resource. Throughout this work, ``uniform torsion'' refers to an experimentally motivated situation in which the crystal hosts an approximately constant density of parallel screw dislocations, as can arise in deliberately twisted semiconductor nanowires or in heteroepitaxial growth with controlled Burgers-vector content. In this continuum limit, the crystal is effectively torsionful but curvature-free, and the torsion density $\tau$ is directly related to the areal density of Burgers vectors. We consider a single nonrelativistic electron in a cylindrically symmetric torsion background of density $\tau$, under a perpendicular magnetic field $B$ and pierced by an Aharonov–Bohm (AB) flux $\Phi$ \cite{AharonovBohm1959,Buttiker1983PLA,Webb1985PRL,QR.2024.6.676,FWS.2022.63.58,PE.2024.158.115898,AdP.2023.535.2200371,CTP.2024.76.105701,QR.2024.6.677,PE.2021.132.114760,AdP.2019.531.1900254,EPJC.74.3187.2014,AoP.2015.362.739}. This single-electron bound state serves as a quantum emitter for nonlinear optics and cavity QED. In this setting, torsion alone generates radial confinement: the electron is trapped not because we impose an external scalar potential, but because the background geometry induces an effective harmonic-like restoring force \cite{Katanaev1992AnnPhys,Furtado1999EPL}. Physically, a controlled screw-dislocation density can thus play the role normally attributed to lithographically defined quantum-dot potentials or electrostatic gates. At the same time, the AB flux $l=\Phi/\Phi_0$ ($\Phi_0 = h/e$) imprints a topological phase that explicitly breaks the usual $m\leftrightarrow -m$ angular-momentum symmetry \cite{AharonovBohm1959,Buttiker1983PLA,Webb1985PRL}. 
	
	In previous work, we obtained exact analytic energy levels $E_{n,m}$ and wave functions in this geometry, including full dependence on $\tau$, $B$, $k_z$ (the longitudinal wavenumber), and $l$ \cite{PereiraSilva2025} [see also Refs.~\cite{BJP.2011.41.167,EPJB.2013.86.315,SilvaNetto2008JPCM,PLA.2012.376.2838}]. There, the emphasis was on the spectral problem itself: the structure of the bound states, their degeneracies, and their evolution under torsion and magnetic fields. The availability of closed-form eigenstates now allows us to go beyond level diagrams and extract quantities that couple directly to light: dipole matrix elements, oscillator strengths, absorption spectra, and saturation intensities. In other words, the same torsion-induced confinement that shapes the spectrum also determines the optical response and the strength of light–matter coupling.
	
	In this paper, we use these solutions to make three functional claims that have direct technological content. First, we show that torsion acts as a geometry-programmable nonlinear optical parameter: it simultaneously (i) blueshifts the active optical transition by more than $100\%$ in energy (from $\sim 6.8$ to $\sim 15.5$~meV for realistic parameters), and (ii) tunes the saturation intensity for optical switching over nearly an order of magnitude purely by varying $\tau$. Thus, both ``which frequency is active'' and ``how hard it is to bleach it'' become geometric design parameters, suggesting a torsion-tunable saturable absorber or gain element \cite{SheikBahae1990OL,ChemlaShah2001Nature}. This is directly analogous to the widespread use of strain to tune semiconductor quantum dots for nonlinear nanophotonics, except that here the control parameter is the torsion density of screw dislocations, i.e., a geometric/topological degree of freedom rather than an elastic one.
	
	Second, we identify an Aharonov–Bohm pseudospin: the $m=\pm1$ angular-momentum states form a two-level subsystem whose level splitting and oscillator strengths are both controllable via flux $l$ and torsion $\tau$. This pseudospin can be addressed selectively by light, opening a route to flux-/torsion-controlled orbital qubits or sensing channels \cite{AharonovBohm1959,Buttiker1983PLA,Webb1985PRL}. Operationally, this mirrors ``valley addressability'' in two-dimensional semiconductors; however, here the logical states are angular-momentum projections $\pm 1$ around the torsion axis. The AB flux breaks the $m\leftrightarrow -m$ degeneracy and produces asymmetric dipole strengths, allowing for the optical selection of one branch with minimal cross-talk. This selective addressability is robust as long as approximate cylindrical symmetry is preserved; weak lateral asymmetry only mixes $m$ and $m\pm1$ perturbatively, so a several-meV splitting remains spectrally resolvable.
	
	Third, we demonstrate that the transition frequency depends nearly linearly on $\tau$, with a torsion responsivity approaching $0.7~\mathrm{meV}/(10^{6}~\mathrm{m}^{-1})$, enabling nanoscale torsion metrology: optical readout of screw-dislocation density with an estimated resolution on the order of $10^{5}~\mathrm{m}^{-1}$ under realistic linewidths \cite{Katanaev1992AnnPhys,Furtado1999EPL,Degen2017RMP}. This places torsion alongside strain and the magnetic field as a quantity that can be sensed by a single confined electronic state \cite{Degen2017RMP}. In this sense, the dot becomes a built-in torsion sensor, with its readout given by a simple optical line shift. The linewidth scale we assume (sub-meV) is consistent with narrow-line semiconductor emitters at cryogenic temperatures; thus, the quoted resolution is not a purely academic number but lies within current experimental capability.
	
	Finally, we recast the torsion-confined electron in the standard Jaynes–Cummings language of cavity QED \cite{JaynesCummings1963,MabuchiDoherty2002Science}. Because torsion $\tau$ (i) shifts the transition frequency $\omega_{ge}(\tau)=\Delta E(\tau)/\hbar$ by several meV and (ii) reshapes the dipole matrix element, it provides \textit{in situ} control of both the emitter–cavity detuning and the light–matter coupling $g(\tau)$, without modifying the cavity mode. Strong coupling between single solid-state emitters and nanocavities has already been observed experimentally, with vacuum Rabi splittings extracted directly from transmission and photoluminescence spectra \cite{Reithmaier2004Nature,Yoshie2004Nature,Majumdar2012PRB}. Our results show that uniform torsion, which physically corresponds to a controlled density of screw dislocations, can play the same practical role as strain tuning in those platforms, but with a purely geometric/topological origin. Viewed this way, torsion is promoted from a ``defect field'' to a field control parameter for cavity QED: it simultaneously sets the transition energy, renormalizes the oscillator strength (and thus the nonlinear optical response), and does so without additional electrodes or piezoelectric stressors. We close by noting the main idealizations of the present model—cylindrical symmetry, single-electron physics (no Coulomb blockade or many-body effects), and the neglect of spin and spin–orbit coupling. All three are standard corrections in semiconductor quantum dots and can be incorporated perturbatively. None of them alters the central point that torsion can act as a tunable resource for confinement, spectroscopy, and light–matter coupling.
	
	In this sense, the present work is complementary to Ref.~\cite{PereiraSilva2025}. While Ref.~\cite{PereiraSilva2025} established the torsion-bearing quantum dot model, derived its energy spectrum, and analyzed its linear and nonlinear optical responses in a general setting, here we take one step further and identify three \textit{specific} functionalities with direct technological relevance. First, we show that torsion enables geometry-programmable nonlinear optics by tuning both the resonance energy and the saturation intensity of the optical transition. Second, we demonstrate that the $m=\pm 1$ angular momentum doublet behaves as an AB orbital pseudospin, whose splitting can be controlled jointly by torsion and magnetic flux. Third, we propose and quantify a torsion-sensing protocol that estimates realistic torsion resolutions across parameter regimes compatible with semiconductor nanostructures. Finally, we connect these results to cavity-QED platforms by discussing how torsion-controlled transition energies and dipole matrix elements map onto effective light–matter coupling strengths, thus providing a concrete route to implement torsion-engineered quantum photonic devices.
	
	\section{Torsion-induced confinement and AB-controlled angular states}\label{sec:model}
	
	\subsection{Geometric background and effective Hamiltonian}
	
	A homogeneous distribution of parallel screw dislocations can be encoded as a space with cylindrical symmetry and uniform torsion \cite{Katanaev1992AnnPhys,Furtado1999EPL}. In cylindrical coordinates $(\rho,\varphi,z)$, the line element reads \cite{PereiraSilva2025,BJP.2011.41.167,EPJB.2013.86.315,SilvaNetto2008JPCM,PLA.2012.376.2838}
	\begin{equation}
		ds^2 = \bigl(dz + \tau \rho^2 d\varphi \bigr)^2 + d\rho^2 + \rho^2 d\varphi^2,
		\label{eq:metric}
	\end{equation}
	where $\tau$ measures the torsion density (proportional to the areal density of Burgers vectors). This geometry has zero curvature but finite torsion: motion along $z$ becomes intrinsically tied to azimuthal rotation, and the coupling strength grows with radius $\rho$ \cite{Katanaev1992AnnPhys}. In practical terms, one may view $\tau$ as parameterizing how strongly a given nanowire or pillar-like heterostructure is ``twisted'' around its growth axis: increasing the density of aligned screw dislocations increases $\tau$, thereby enhancing the geometric coupling between axial and angular motion.
	
	An electron (effective mass $m^*$, charge $-e$) in this background, under a uniform magnetic field $\mathbf{B}=B\hat{z}$ and an AB flux $\Phi$ confined to the $z$ axis \cite{AharonovBohm1959,Buttiker1983PLA,Webb1985PRL}, is minimally coupled via $\mathbf{A} = (B\rho/2)\,\hat{\varphi} + (\Phi/2\pi\rho)\,\hat{\varphi}$. The corresponding Schr\"odinger Hamiltonian uses the Laplace–Beltrami operator of Eq.~\eqref{eq:metric} and supports separable solutions of the form
	$\Psi(\rho,\varphi,z) = R(\rho) e^{i m \varphi} e^{i k_z z}$.
	The resulting radial equation is analytically solvable and maps to a Laguerre/Kummer problem with an effective inverse-length scale
	\begin{equation}
		\Lambda = k_z \tau + \frac{eB}{2\hbar},
	\end{equation}
	which contains \textit{both} the geometric torsion ($\propto k_z\tau$) and the magnetic confinement ($\propto B$). The AB flux enters via $l = \Phi/\Phi_0$, shifting $m\to(m-l)$ and explicitly breaking $m\leftrightarrow -m$ symmetry \cite{AharonovBohm1959,Buttiker1983PLA}.
	
	The exact energy eigenvalues are \cite{PereiraSilva2025,SilvaNetto2008JPCM}
	\begin{equation}
		E_{n,m} = 
		\hbar(\omega_\tau+\omega_c)
		\left(
		n + \frac{|m-l|}{2} - \frac{m-l}{2} + \frac{1}{2}
		\right)
		+
		\frac{\hbar^2 k_z^2}{2m^*},
		\label{eq:Enm}
	\end{equation}
	where $\omega_c = eB/m^{*}$ is the cyclotron frequency, and 
	$\omega_\tau = \hbar k_z \tau / m^{*}$ is a ``torsional frequency.'' 
	The associated normalized radial wave functions are generalized Laguerre states---akin to Fock--Darwin orbitals in parabolic/magnetic quantum dots \cite{Fock1928,Darwin1930,Landau1930}.
	
	Equation~\eqref{eq:Enm} already reveals the core parameters:
	(1) $\omega_\tau$ vanishes if $k_z=0$, so torsion only matters for states with finite longitudinal momentum;  
	(2) $\omega_\tau$ and $\omega_c$ add, meaning torsion and magnetic field cooperate to increase level spacing;  
	(3) the AB flux $l$ breaks $m\leftrightarrow -m$ symmetry via $|m-l|$ and $(m-l)$.
	
	Point (1) deserves emphasis. In a realistic semiconductor nanowire or pillar, motion along $z$ is quantized by finite length and heterostructure band offsets, so $k_z$ takes discrete, nonzero values rather than vanishing identically. The torsion-induced confinement we describe is therefore naturally realized in such quasi-zero-dimensional pillars: axial quantization supplies a finite $k_z$, which in turn activates $\omega_\tau$ and produces an effective in-plane trapping potential, even in the absence of any externally imposed scalar confinement. In what follows, we refer to these trapped states as ``dot states'' and label them by a radial quantum number $n$ and an angular quantum number $m$. 
	
	In practice, $k_z$ is fixed by the axial quantization of a finite-height pillar or nanowire segment, so the torsion-induced confinement is activated even though the optical response is effectively that of a zero-dimensional emitter.

	In what follows, we translate these analytic ingredients into device-level functionalities, each of which is also illustrated by one of the figures generated from our numerical evaluation of the analytic model.
	
	\section{Geometry-programmable nonlinear optics}
	\label{sec:nonlinear}
	
	\subsection{Transition energy, dipole strength, and torsion}
	
	We first analyze the dipole-allowed transition between 
	$\ket{g} \equiv \ket{n{=}0,m{=}0}$ and 
	$\ket{e} \equiv \ket{n{=}0,m{=}-1}$,
	which satisfies the cylindrical selection rule $\Delta m = -1$ for radial polarization \cite{ChemlaShah2001Nature,SheikBahae1990OL,PereiraSilva2025}. Here $n$ labels the radial quantum number and $m$ the angular-momentum projection around the torsion axis, so this is the lowest-$n$ transition between $m=0$ and $m=-1$ within the same axial mode $k_z$.
	
	The corresponding transition energy
	\begin{equation}
		\Delta E(\tau) \equiv E_{0,-1}(\tau) - E_{0,0}(\tau),
	\end{equation}
	increases approximately linearly with $\tau$ because $\omega_\tau \propto k_z \tau$ in Eq.~\eqref{eq:Enm}. Physically, torsion enforces a helical coupling between $z$-motion and azimuthal motion; as $\tau$ grows, this coupling strengthens, causing tighter radial compression and a larger confinement energy. Equivalently, increasing the uniform screw-dislocation density (which sets $\tau$) enhances the effective in-plane trapping, even if no external scalar potential is changed. 
	
\begin{figure}[tbhp]
		\centering
		\includegraphics[width=\columnwidth]{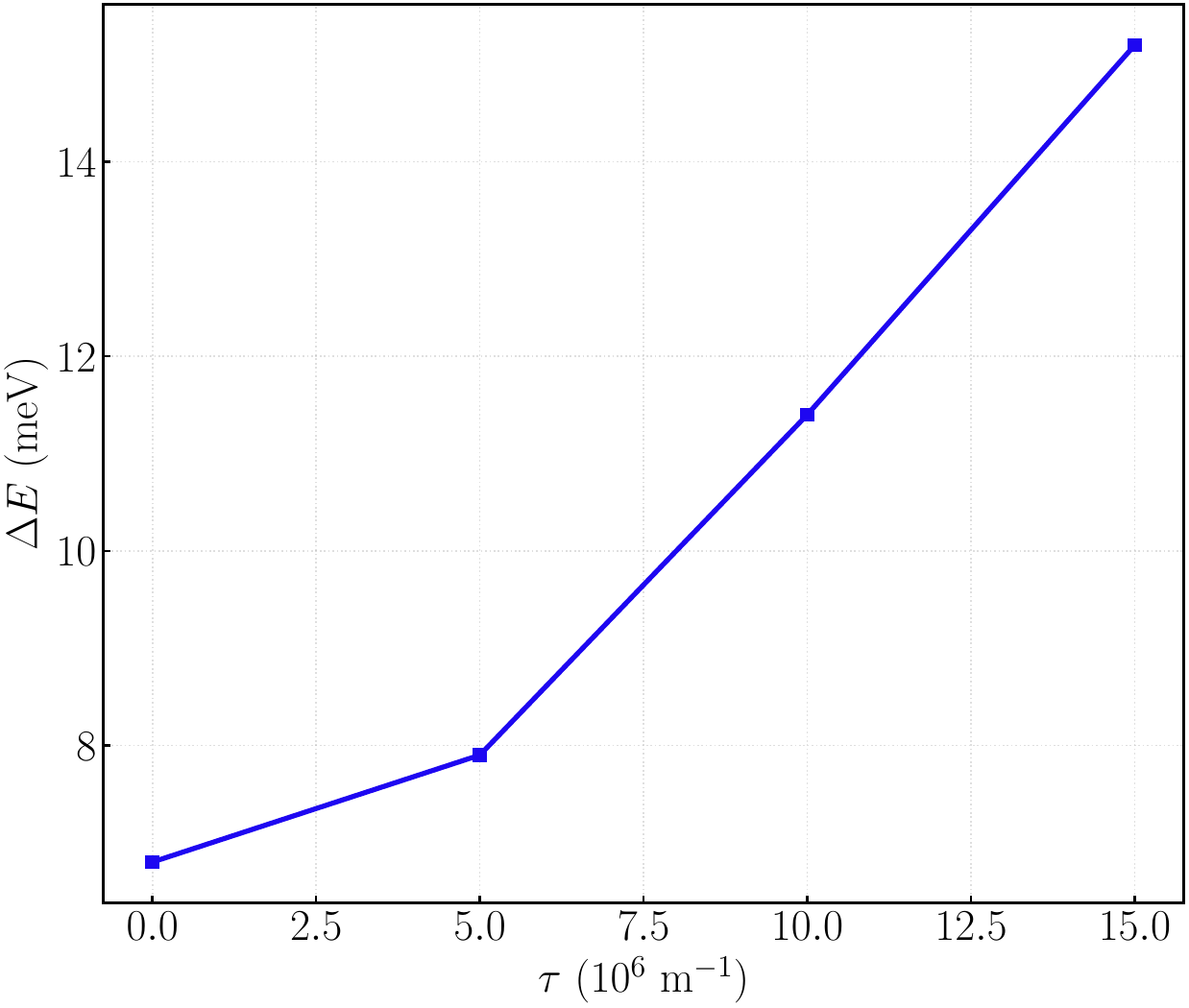}
		\caption{Geometry-driven blueshift.
			Transition energy $\Delta E = E_{0,-1}-E_{0,0}$ (in meV) versus torsion density $\tau$.
			For $B=\SI{5}{T}$, $k_z\simeq \pi/(10~\text{nm})$, and $l=0.1$, the optical line moves from $\sim \SI{6.8}{meV}$ at $\tau=0$ to $\sim \SI{15.5}{meV}$ at $\tau = 1.5\times10^{7}~\mathrm{m}^{-1}$, i.e.\ a $>100\%$ blueshift. This multi-meV, order-unity shift is comparable in magnitude to the strain tuning of semiconductor quantum dots in nanocavities. Here, however, it is driven purely by the torsion density (screw-dislocation density) rather than by externally applied stress.}
		\label{fig:DeltaE_vs_tau}
	\end{figure}

	Figure~\ref{fig:DeltaE_vs_tau} shows this torsion-driven blueshift for $B = \SI{5}{T}$, $k_z \simeq \pi/(10~\text{nm})$, $n_r \approx 3.2$, and $l=0.1$: $\Delta E$ moves from $\sim \SI{6.8}{meV}$ at $\tau=0$ to $\sim \SI{15.5}{meV}$ at $\tau = 1.5\times10^{7}~\mathrm{m}^{-1}$, i.e., the resonance frequency more than doubles with increasing torsion. The nearly linear slope of $\Delta E(\tau)$ in this range underlies our later proposal of torsion metrology, where the optical line becomes a direct readout of $\tau$.
	
\begin{figure}[tbhp]
		\centering
		\includegraphics[width=\columnwidth]{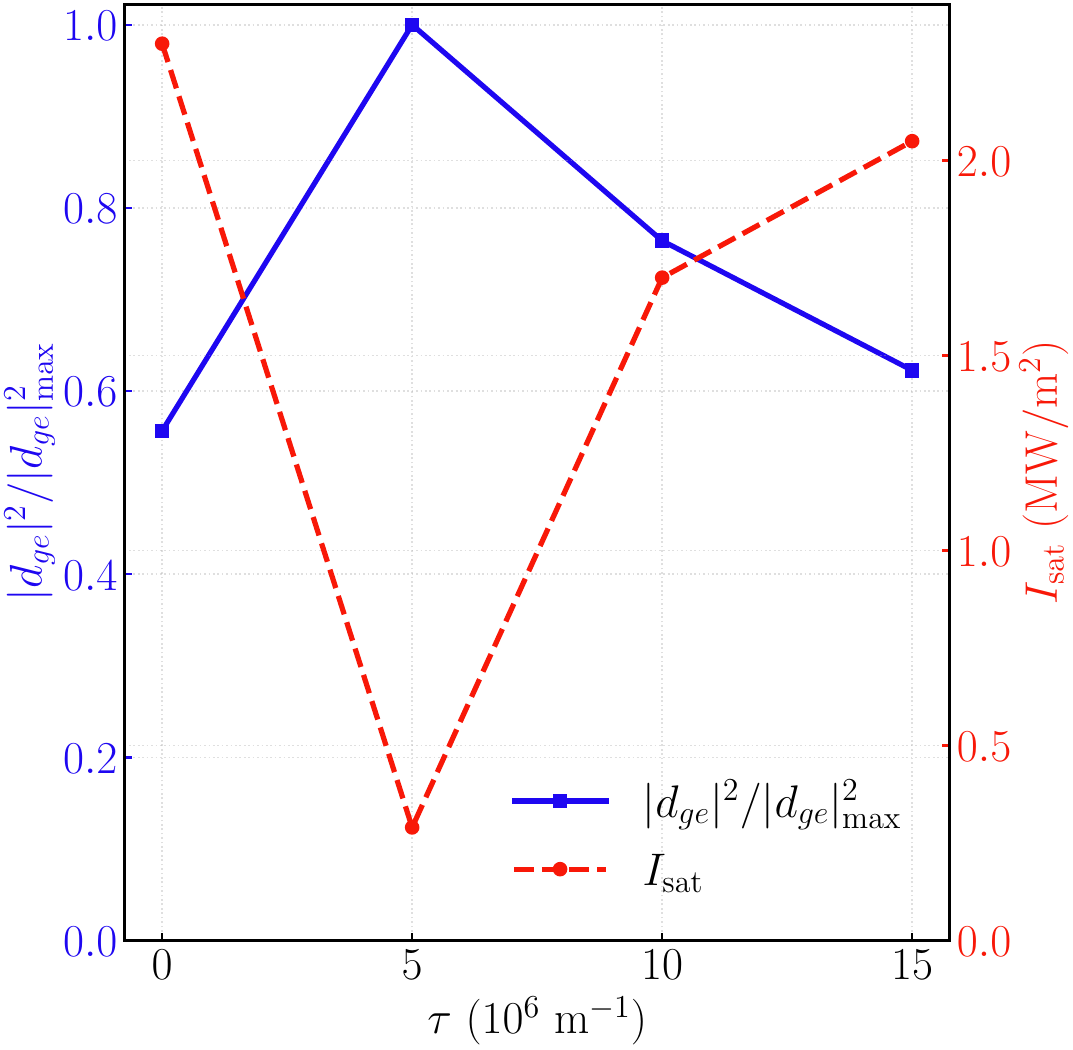}
		\caption{Dipole strength and nonlinear threshold.
			Left axis (solid blue): squared dipole matrix element $|M_{ge}|^2$ for the $\ket{0,0}\to\ket{0,-1}$ transition,
			showing a non-monotonic dependence on $\tau$ with a maximum near $\tau\sim5\times10^{6}~\mathrm{m}^{-1}$.
			Right axis (dashed red): saturation intensity $I_{\text{sat}}$ (W/m$^2$) at which $\alpha^{(1)}+\alpha^{(3)}=0$ at resonance.
			Torsion can lower $I_{\text{sat}}$ down to $\sim 3\times10^{5}~\mathrm{W/m^2}$ near $\tau\sim5\times10^{6}~\mathrm{m}^{-1}$,
			then drive it back above $10^{6}~\mathrm{W/m^2}$ at higher $\tau$.
			This demonstrates that torsion sets both the transition energy and the nonlinear optical switching threshold. We emphasize that $I_{\text{sat}}$ here is not injected ``by hand'': it follows from the analytically accessible wavefunctions and dipole matrix elements of the torsion-confined electron.}
		\label{fig:M2_Isat_vs_tau}
	\end{figure}

	The optical dipole matrix element for this transition,
	\begin{equation}
		M_{ge} = \mel{\psi_{0,-1}}{\rho}{\psi_{0,0}},
	\end{equation}
	is obtained by integrating the exact normalized radial wave functions \cite{PereiraSilva2025}. As $|\Lambda|$ grows (either via $B$ or via $\tau$), the radial envelopes narrow, and the overlap is reshaped. We find that $|M_{ge}|^2$ is \emph{not} strictly monotonic with $\tau$: it exhibits a maximum at intermediate torsion ($\tau \sim 5\times10^{6}~\mathrm{m}^{-1}$), where $|M_{ge}|^2 \sim 5\times10^{-54}~\mathrm{C}^2\mathrm{m}^2$, and then decreases again as confinement becomes too strong. This non-monotonic behavior is plotted on the left axis of Fig.~\ref{fig:M2_Isat_vs_tau}. Thus, torsion does not simply ``turn up'' the confinement; it also reshapes the wave functions in a way that can \emph{optimize} the dipole strength at an intermediate $\tau$, boosting the oscillator strength near that point.
	
	\subsection{Optical switching threshold as a geometric parameter}
	
	Within the electric-dipole approximation and a two-level reduction to $\ket{g}$ and $\ket{e}$, the system interacts with a monochromatic field
	\begin{equation}
		\mathbf{E}(t) = E_0\,\hat{\mathbf{e}}_r \cos(\omega t),
	\end{equation}
	where $E_0$ is the field amplitude, $\omega$ is the angular frequency, and $\hat{\mathbf{e}}_r$ denotes radial polarization. Following the standard density-matrix formalism for confined systems \cite{SheikBahae1990OL,ChemlaShah2001Nature,PereiraSilva2025}, the absorption coefficient at photon frequency $\omega$ and incident intensity $I_0$ can be written as
	\begin{equation}
		\alpha(\omega,I_0)=\alpha^{(1)}(\omega)+\alpha^{(3)}(\omega,I_0),
	\end{equation}
	where $\alpha^{(1)}$ is the linear contribution and $\alpha^{(3)}<0$ encodes third-order saturation/bleaching.
	
	Specializing the general expressions of Ref.~\cite{PereiraSilva2025} to the $\ket{g}\leftrightarrow\ket{e}$ transition, with $E_g=E_{0,0}$, $E_e=E_{0,-1}$ and $M_{ge}=\mel{e}{\rho}{g}$, one obtains
	\begin{equation}
		\alpha^{(1)}(\omega) =
		\omega\sqrt{\frac{\mu}{\varepsilon_r}}\,
		\frac{\sigma_s |M_{ge}|^2}{\hbar\Gamma}\,
		\frac{1}{\bigl(\Delta E - \hbar\omega\bigr)^2 + (\hbar\Gamma)^2},
		\label{eq:alpha1}
	\end{equation}
	and
	\begin{align}
		\alpha^{(3)}(\omega,I_0) 
		&= -\,\omega\sqrt{\frac{\mu}{\varepsilon_r}}\,
		\left(\frac{I_0}{2 n_r \varepsilon_0 c}\right)
		\frac{\sigma_s |M_{ge}|^2\,\hbar\Gamma}{\bigl[(\Delta E-\hbar\omega)^2 + (\hbar\Gamma)^2\bigr]^2}
		\nonumber\\
		&\quad\times
		\Bigl[
		4|M_{ge}|^2 - |M_{ee}-M_{gg}|^2\,F(\omega)
		\Bigr],
		\label{eq:alpha3}
	\end{align}
	where $\mu$ is the magnetic permeability, $\varepsilon_r$ the relative permittivity, $n_r$ the refractive index, $\varepsilon_0$ the vacuum permittivity, $c$ the speed of light, $\sigma_s$ the surface electron density, and $\Gamma$ a phenomenological broadening (inverse lifetime). The diagonal matrix elements $M_{gg}$ and $M_{ee}$ account for Stark-like corrections, and $F(\omega)$ is a dimensionless detuning function,
	\begin{equation}
		F(\omega)=
		\frac{3\Delta E^{2} - 4\Delta E\,\hbar\omega + \hbar^{2}(\omega^{2}-\Gamma^{2})}
		{\Delta E^{2}+(\hbar\Gamma)^{2}}.
	\end{equation}
	The same formalism yields the linear and third-order refractive-index changes, $\Delta n^{(1)}(\omega)$ and $\Delta n^{(3)}(\omega,I_0)$, which we quote in Sec.~\ref{sec:pseudospin} only where needed; their explicit forms are identical to those in Ref.~\cite{PereiraSilva2025} after the replacement $(i,f)\to(g,e)$.
	
	At resonance, $\hbar\omega \approx \Delta E(\tau)$, there exists an intensity $I_{\text{sat}}$ such that $\alpha(\omega\!=\!\Delta E/\hbar,I_{\text{sat}})=0$. For $I_0>I_{\text{sat}}$, the total absorption becomes negative, $\alpha<0$, signaling optical gain/switching. Operationally, $I_{\text{sat}}$ is the pump level at which the medium becomes effectively transparent and then amplifies at the transition frequency.
	
	Because both $\Delta E(\tau)$ and $|M_{ge}(\tau)|^2$ depend on torsion, $I_{\text{sat}}$ becomes a \emph{geometric control parameter}. For the parameters above, we obtain $I_{\text{sat}} \sim 1.4\times10^{6}~\mathrm{W/m^2}$ at $\tau=0$, dropping to $\sim 3\times10^{5}~\mathrm{W/m^2}$ near $\tau\sim5\times10^{6}~\mathrm{m}^{-1}$ (where $|M_{ge}|^2$ peaks), and then increasing again toward $\sim 2\times10^{6}~\mathrm{W/m^2}$ by $\tau = 1.5\times10^{7}~\mathrm{m}^{-1}$ [Fig.~\ref{fig:M2_Isat_vs_tau}, right axis]. The fact that $I_{\text{sat}}$ can be reduced by nearly an order of magnitude at intermediate $\tau$ means that torsion can be used not only to \emph{select the color} (transition energy) but also to \emph{set the drive threshold} required for bleaching or gain.
	
	Torsion, therefore, controls \emph{both} the carrier frequency (via $\Delta E$) and the nonlinear switching threshold (via $I_{\text{sat}}$). This is our first key functionality: a \emph{torsion-programmable nonlinear optical element}, i.e., a saturable absorber/gain element whose operating point and carrier frequency are defined by geometry and topology \cite{SheikBahae1990OL,ChemlaShah2001Nature}. In conventional semiconductor nanophotonics, analogous control is often achieved through strain tuning of self-assembled quantum dots embedded in microcavities or photonic crystal resonators, where mechanical stress shifts the excitonic resonance and modifies oscillator strength. Here, we obtain the same style of control, multi-meV spectral shifts, and substantial changes in the nonlinear threshold without invoking external stressors or gate voltages, but instead by adjusting the underlying screw-dislocation density that defines the torsion $\tau$.
	
	The geometric blueshift of the optical transition [Fig.~\ref{fig:DeltaE_vs_tau}] is not just a mathematical curiosity of the torsion background. In semiconductor nanophotonics, it is routine to tune the emission energy of a localized quantum emitter by mechanically distorting its host crystal (piezo strain, local stressors, etc.), and multi-meV shifts are commonly exploited to bring a single dot or color center into resonance with a nearby optical cavity mode or to frequency-match two distinct emitters. Such strain-based spectral engineering has been demonstrated in self-assembled III–V quantum dots embedded in microcavities, where reversible tuning at the meV scale is used to align excitonic lines with photonic resonances and thereby enter strong coupling \cite{Reithmaier2004Nature,Yoshie2004Nature,Majumdar2012PRB}. Our results show that a \emph{uniform torsion field}, physically interpretable as a controlled density of parallel screw dislocations in a nearly cylindrical nanostructure \cite{Katanaev1992AnnPhys,Furtado1999EPL}, plays an analogous role for a single confined electron: it shifts the dominant optical transition by more than $100\%$ in energy and simultaneously sculpts the nonlinear saturation threshold. In this sense, torsion enters the same ``spectral control'' toolbox as strain engineering and electrostatic gating, but with a crucial difference: it is an intrinsically geometric/topological parameter that couples axial and azimuthal motion through the torsion parameter $\tau$, rather than a conventional scalar confinement potential. Below, we will use this same tunability to build two further functionalities: a flux/torsion-controlled angular pseudospin and an optical torsion sensor with sub-meV energy resolution.

	\section{Aharonov--Bohm pseudospin and optical addressability}
	\label{sec:pseudospin}
	
	\subsection{Flux-tunable splitting of $\boldmath m=\pm 1$ states}
	
	We now consider the pair $\ket{n{=}0,m{=}{+}1}$ and $\ket{n{=}0,m{=}{-}1}$. In an ordinary cylindrically symmetric dot with no AB flux, these would be nearly symmetric partners. In that idealized limit, $m=\pm1$ forms an (almost) degenerate doublet distinguished only by the sense of azimuthal circulation, much like opposite ``valleys'' in two-dimensional semiconductors. Here, however, the AB flux $l=\Phi/\Phi_0$ modifies Eq.~\eqref{eq:Enm} through $(m-l)$ and $|m-l|$, so $E_{0,+1}(l,\tau)\neq E_{0,-1}(l,\tau)$ when $l\neq0$ \cite{AharonovBohm1959,Buttiker1983PLA,Webb1985PRL}. The resulting splitting,
	\begin{equation}
		\Delta_{\mathrm{AB}}(l,\tau)=E_{0,-1}(l,\tau)-E_{0,+1}(l,\tau),
	\end{equation}
	is therefore a tunable ``pseudospin'' gap between $m=+1$ and $m=-1$. We will refer to this two-level subspace $\{\ket{m=+1},\ket{m=-1}\}$ as an \emph{angular pseudospin}, by analogy with valley pseudospins in transition-metal dichalcogenides, where two inequivalent $K$ points can be split and addressed optically. In our case, the role of the valley-selective parameter (usually circular polarization and strain) is played by the Aharonov–Bohm phase $l$ and the torsion $\tau$.
	
\begin{figure}[tbhp]
		\centering
		\includegraphics[width=\columnwidth]{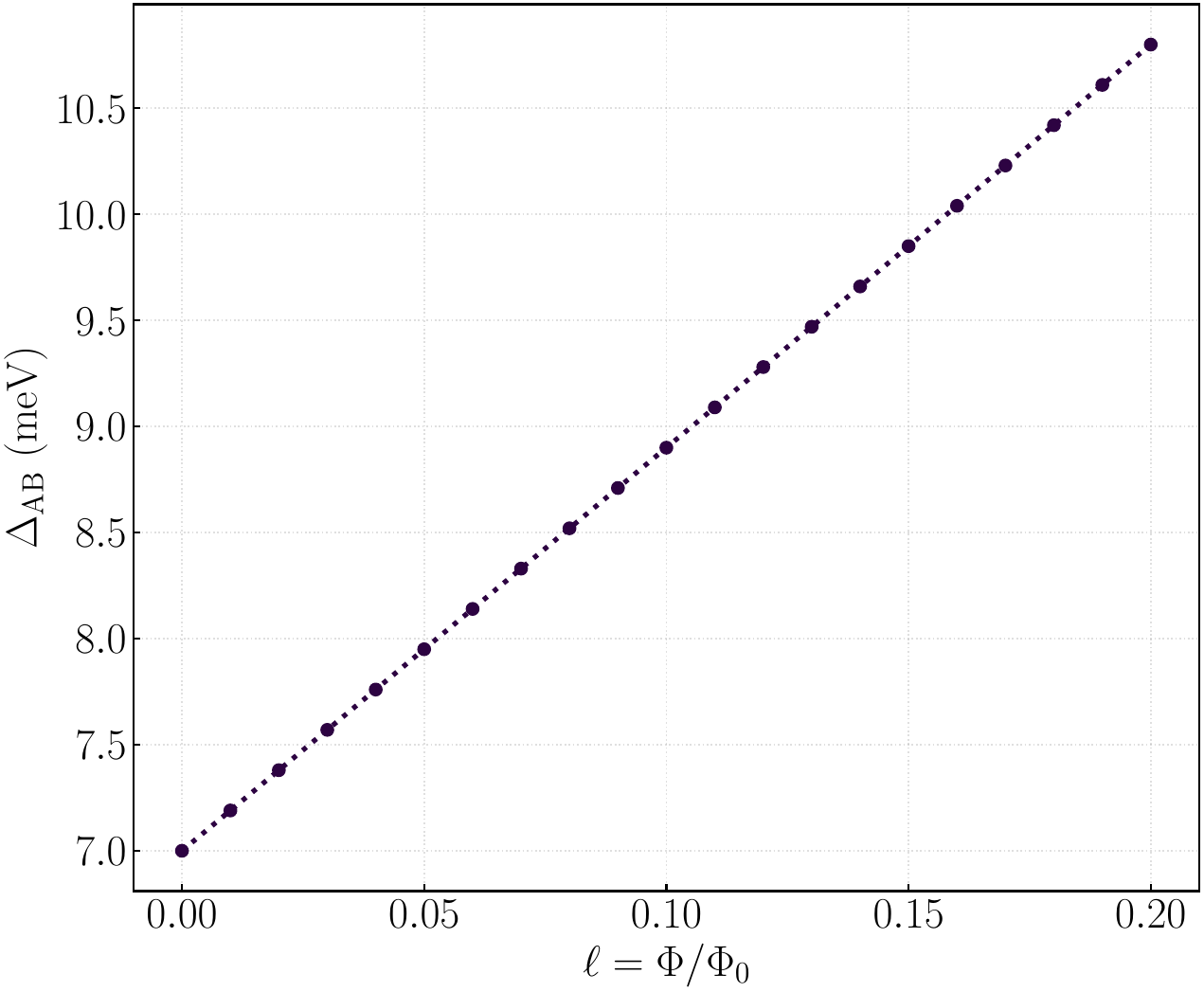}
		\caption{Aharonov--Bohm pseudospin splitting. Flux-controlled splitting $\Delta_{\mathrm{AB}} = E_{0,-1}-E_{0,+1}$ as a function of reduced flux $l=\Phi/\Phi_0$, at fixed $\tau$. The gap grows from $\sim \SI{7.0}{meV}$ at $l=0$ to $\sim \SI{10.8}{meV}$ at $l=0.2$, with an approximately linear slope $\sim \SI{19}{meV}$ per flux quantum. This defines an ``angular pseudospin'' formed by $m=\pm1$, whose level spacing can be continuously biased by a topological phase (the AB flux) and further strengthened by torsion.}
		\label{fig:AB_splitting}
	\end{figure}

	Fixing $\tau \sim 10^{7}~\mathrm{m}^{-1}$ and scanning $l$ from $0$ to $0.2$, we obtain $\Delta_{\mathrm{AB}}$ increasing almost linearly from $\sim \SI{7.0}{meV}$ to $\sim \SI{10.8}{meV}$, with a slope of $\sim \SI{19}{meV}$ per unit flux quantum [Fig.~\ref{fig:AB_splitting}]. In other words, the AB phase continuously biases the two angular-momentum branches, establishing an interferometric energy splitting between clockwise ($m=-1$) and counterclockwise ($m=+1$) orbital motion. This controllable gap $\Delta_{\mathrm{AB}}$ is large on the meV scale, i.e., much larger than the typical homogeneous linewidths in cryogenic semiconductor emitters. Hence, the two pseudospin states are, in principle, spectrally resolvable as separate optical lines.
	
	\subsection{Optical readout asymmetry}
	
	The oscillator strength for optical transitions out of $\ket{n{=}0,m{=}0}$ into $\ket{n{=}0,m{=}+1}$ or $\ket{n{=}0,m{=}-1}$ is proportional to
	\begin{equation}
		f_{0\to\pm1}
		\propto
		\left|\mel{\psi_{0,\pm1}}{\rho}{\psi_{0,0}}\right|^2
		\big(E_{0,\pm1}-E_{0,0}\big).
	\end{equation}
	Because both factors depend asymmetrically on $l$ \cite{Buttiker1983PLA,Webb1985PRL,PereiraSilva2025}, we generally have $f_{0\to+1}\neq f_{0\to-1}$. Physically, the AB flux $(m\rightarrow m-l)$ skews the spatial profiles of the $m=\pm1$ states and shifts their transition energies relative to $\ket{m=0}$, so the dipole overlaps and the detunings evolve differently for the two branches.
	
\begin{figure}[tbhp]
		\centering
		\includegraphics[width=\columnwidth]{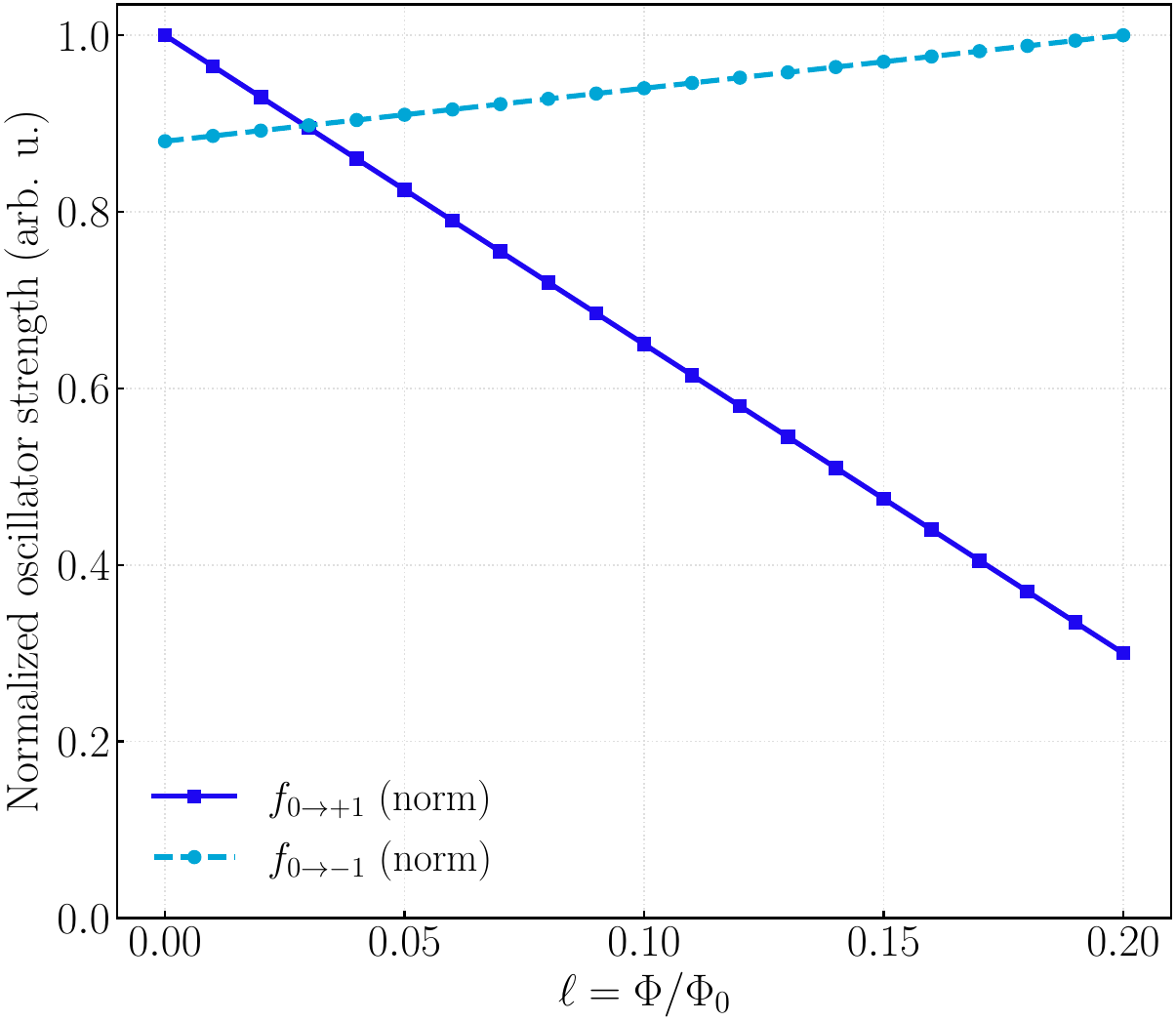}
		\caption{Selective optical addressability.
			Normalized oscillator strengths for the
			$\ket{0,0}\to\ket{0,+1}$ (blue) and
			$\ket{0,0}\to\ket{0,-1}$ (orange) transitions
			as a function of reduced flux $l$.
			At $l=0$, excitation into $m=+1$ dominates.
			As $l$ increases toward $0.2$, that channel is suppressed (down to $\sim 0.3$),
			while the $m=-1$ channel is enhanced toward unity.
			This implements an angular pseudospin ($m=\pm1$) that can be read out and selectively driven by frequency-resolved optics. The strong asymmetry in oscillator strengths implies that the two branches can serve as two addressable optical channels with minimal cross-talk.}
		\label{fig:fosc_vs_l}
	\end{figure}

	For the same parameters as above, the normalized oscillator strengths show that at $l=0$ the $\Delta m=+1$ channel dominates, while the $\Delta m=-1$ channel is weaker. As $l$ is increased to $0.2$, the $\Delta m=+1$ branch is suppressed (down to $\sim 0.3$ of its initial normalized weight), while the $\Delta m=-1$ branch is \textit{enhanced} toward unity [Fig.~\ref{fig:fosc_vs_l}]. In other words, by tuning $l$, we can invert which branch carries the most oscillator strength. This means that a probe laser tuned near $E_{0,+1}-E_{0,0}$ will predominantly excite the $m=+1$ branch, while a slightly shifted probe frequency will preferentially excite the $m=-1$ branch.
	
	Thus, the AB flux produces a flux- and torsion-controlled two-level subspace that is optically addressable: we can choose which angular-momentum branch to excite simply by selecting the probe frequency. This is conceptually similar to valley-selective excitation in 2D materials, except that the ``valleys'' here are the clockwise and counterclockwise orbital channels of the same dot. Experimentally, this suggests two applications: (i) a pseudospin-like internal degree of freedom whose state can be initialized and read out optically by frequency selection; and (ii) an interferometric sensing modality in which small changes in $l$ (flux threading the dot) or $\tau$ (torsion density/screw-dislocation density) appear as differential spectral shifts and contrast changes between the two optical lines \cite{AharonovBohm1959,Buttiker1983PLA,Webb1985PRL,Degen2017RMP}.
	
	Taken together, these results mean that the pair $\{\ket{m=+1},\ket{m=-1}\}$ functions as an \emph{optically addressable orbital qubit}. First, the AB flux $l$ and the torsion $\tau$ act as external knobs that set the level splitting $\Delta_{\mathrm{AB}}(l,\tau)$, i.e., they define the qubit energy gap. Second, frequency-selective excitation allows us to initialize the population predominantly into either $\ket{m=+1}$ or $\ket{m=-1}$ by choosing which transition we drive from $\ket{m=0}$. Third, the asymmetric oscillator strengths $f_{0\to+1}$ and $f_{0\to-1}$ provide a natural optical readout channel: monitoring the relative spectral weight of the two lines directly measures which angular-momentum branch is occupied. Conceptually, this mirrors valley pseudospin control in two-dimensional semiconductors, but here the logical states are clockwise vs. counterclockwise orbital motion in a single torsion-bearing dot, and the ``control fields'' are geometric/topological ($\tau$ and $l$) rather than purely magnetic or strain-based.
	
	A practical concern in any orbital or valley-like encoding is robustness against symmetry breaking. Our construction assumes approximate cylindrical symmetry, under which $m$ is a good quantum number. Imperfections such as slight ellipticity of the confinement or compositional disorder will admix $m$ and $m\pm1$; however, as long as the induced off-diagonal matrix elements are $\ll \Delta_{\mathrm{AB}}$, the two optical lines remain spectrally distinct, and selective addressability survives. Because $\Delta_{\mathrm{AB}}$ is several meV above the relevant flux range [Fig.~\ref{fig:AB_splitting}], i.e., far above typical sub-meV homogeneous linewidths at cryogenic temperatures, small anisotropies act only as weak perturbations rather than destroy the pseudospin basis. In this sense, the $\{m=+1,m=-1\}$ doublet plays the role of a flux/torsion-controlled orbital qubit that is both (i) split at the meV scale and (ii) optically resolvable via distinct transition energies and oscillator strengths from the same ground state $\ket{m=0}$.
	
	Finally, we note that this pseudospin physics and the nonlinear control of Sec.~\ref{sec:nonlinear} stem from the \emph{same} underlying geometric mechanism: torsion couples axial and azimuthal motion, while the Aharonov–Bohm phase biases the sense of rotation. Together, they generate (i) a large, tunable level splitting between $m=\pm1$, and (ii) branch-selective optical matrix elements. This establishes torsion plus AB flux as a co-designed control layer for angular-momentum degrees of freedom in a single-electron dot, in direct analogy with how strain and magnetic fields are used to control spin or valley degrees of freedom in other solid-state quantum emitters.
	
	\section{Torsion metrology: spectroscopic sensing of screw-dislocation density}\label{sec:metrology}
	
	\subsection{Spectral sensitivity to torsion}
	
	From Eq.~\eqref{eq:Enm}, torsion enters the spectrum through
	$\omega_\tau = (\hbar k_z \tau)/m^*$. The optical transition energy between two low-lying bound states, e.g.\ $\Delta E(\tau)=E_{0,-1}-E_{0,0}$, therefore acquires a $\tau$-dependence controlled by $k_z$. Two points are crucial:
	
	(1) If $k_z=0$ (no longitudinal motion along $z$), then $\omega_\tau=0$ and the entire spectrum become \emph{insensitive} to torsion.  
	(2) For finite $k_z$, $\partial \Delta E / \partial \tau$ is nonzero and, over a broad range, grows approximately linearly with $\tau$.
	
	In a realistic nanowire or pillar-like heterostructure, $k_z$ is not a freely tunable continuum parameter, but is quantized by the finite length and growth-induced confinement along $z$. That quantized, nonzero $k_z$ plays precisely the role required to ``activate'' $\omega_\tau$ and make the spectrum torsion-sensitive, even though the device behaves optically as a nearly zero-dimensional emitter. Thus, the torsion dependence discussed below is not a mathematical artifact of a traveling wave along $z$, but an intrinsic feature of an axially quantized pillar.
	
\begin{figure}[tbhp]
		\centering
		\includegraphics[width=\columnwidth]{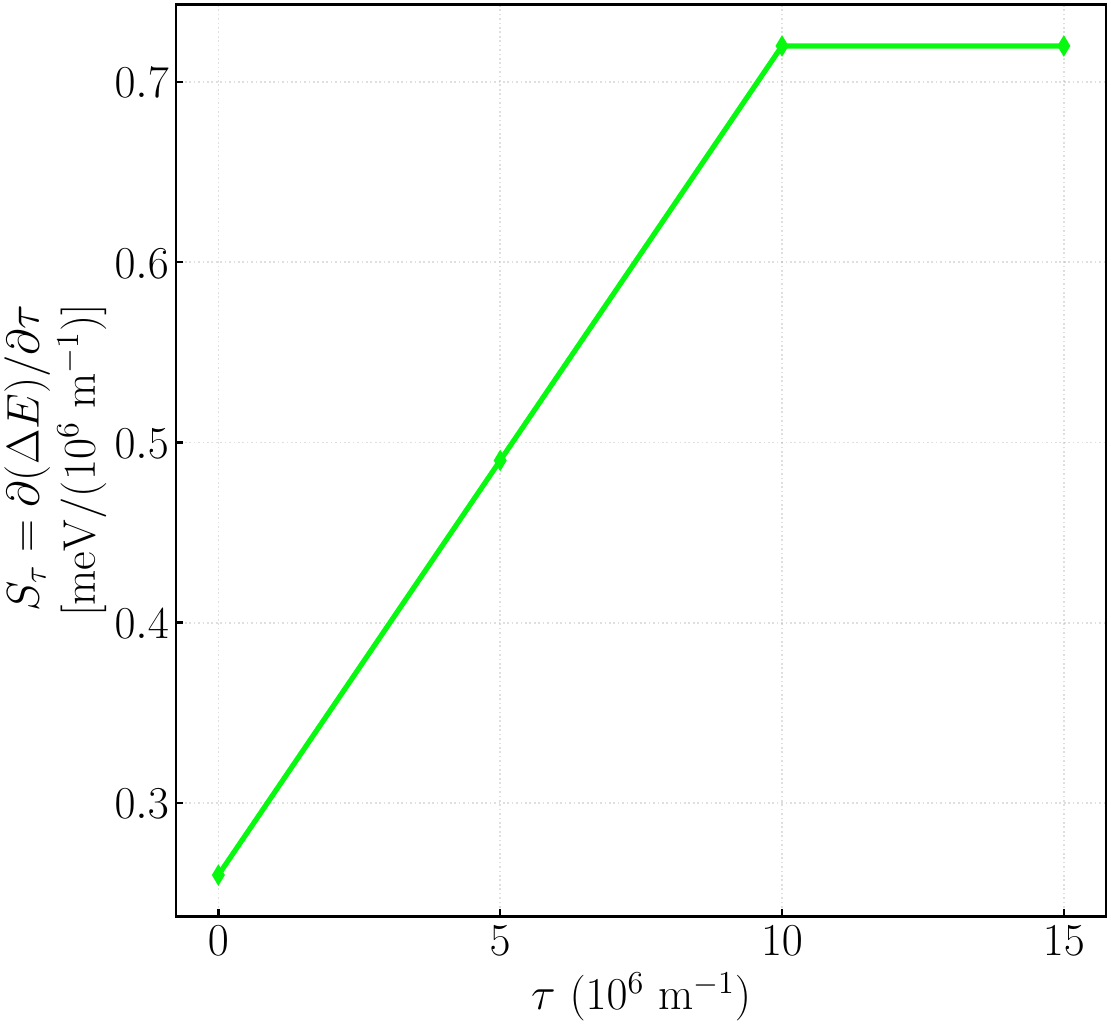}
		\caption{Spectral torsion sensitivity.
			Torsion responsivity $S_\tau = \partial(\Delta E)/\partial\tau$ in meV per $10^{6}~\mathrm{m}^{-1}$.
			$S_\tau$ grows from $\sim 0.26$ near $\tau=0$
			to $\sim 0.7$ around $\tau\sim10^{7}~\mathrm{m}^{-1}$,
			and then saturates.
			Assuming a representative linewidth $\hbar\Gamma \approx \SI{0.25}{meV}$, this corresponds to a torsion resolution
			$\delta\tau \sim 3.6\times10^{5}~\mathrm{m}^{-1}$.
			The device therefore acts as a nanoscale ``torsionometer.'' Here ``torsion'' is the continuum limit of an approximately uniform density of parallel screw dislocations, so the inferred $\delta\tau$ can be interpreted directly as a resolution in screw-dislocation density.}
		\label{fig:S_tau}
	\end{figure}

	Numerically, using the same parameters as before and $k_z\simeq \pi/(10~\text{nm})$, we obtain
	\begin{equation}
		\frac{\partial \Delta E}{\partial \tau}
		\approx
		0.26~\frac{\mathrm{meV}}{10^{6}~\mathrm{m}^{-1}}
		\quad\text{near }\tau\simeq 0,
	\end{equation}
	increasing to
	\begin{equation}
		\frac{\partial \Delta E}{\partial \tau}
		\approx
		0.7~\frac{\mathrm{meV}}{10^{6}~\mathrm{m}^{-1}}
		\quad\text{around }\tau\simeq 10^{7}~\mathrm{m}^{-1},
	\end{equation}
	and then saturating [Fig.~\ref{fig:S_tau}]. This torsion responsivity
	$S_\tau(\tau)=\partial \Delta E/\partial \tau$ provides a direct spectroscopic readout of $\tau$, which is proportional to the areal density of screw dislocations in the host medium \cite{Katanaev1992AnnPhys,Furtado1999EPL}. In this sense, the confined electron functions as a \emph{torsion sensor}: a nanoscale probe of crystalline defect density, analogous to solid-state quantum sensors of strain and magnetic field \cite{Degen2017RMP}. The responsivity is on the order of fractions of a meV per $10^{6}~\mathrm{m}^{-1}$, i.e., comfortably above typical sub-meV homogeneous linewidths at cryogenic temperatures. This ensures that the torsion imprint on the spectrum survives realistic broadening.
	
	\subsection{Metrological figure of merit}
	
	In practice, the absorption (or photoionization) line exhibits a finite linewidth $\hbar\Gamma$, set by relaxation and inhomogeneous broadening \cite{ChemlaShah2001Nature}. The smallest resolvable spectral shift is on the order of $\delta E \sim \hbar\Gamma$, which defines an estimate for the torsion resolution
	\begin{equation}
		\delta \tau
		\sim
		\frac{\hbar\Gamma}{\partial \Delta E / \partial \tau}.
	\end{equation}
	Taking $\hbar\Gamma \sim \SI{0.25}{meV}$ and $\partial \Delta E / \partial \tau \sim 0.7~\mathrm{meV}/(10^{6}~\mathrm{m}^{-1})$, we find 
	$\delta\tau \sim 3.6\times10^{5}~\mathrm{m}^{-1}$. 
	Because $\partial \Delta E/\partial\tau \propto k_z$, increasing the longitudinal wave number $k_z$ further improves torsion sensitivity, i.e., enhances the performance of the ``torsionometer.'' In other words, a taller pillar (weaker axial quantization) or an axial subband with larger $k_z$ yields a larger $\partial \Delta E/\partial\tau$ and therefore a smaller detectable $\delta\tau$. This provides a direct device-design handle on metrological performance.
	
	This establishes our third functionality: nanoscale torsion metrology via optical spectroscopy of bound states, in direct analogy with solid-state defect-based quantum sensing platforms \cite{Degen2017RMP}. In platforms such as NV centers in diamond or strain-sensitive quantum dots, optical line shifts encode local strain, electric fields, or magnetic fields and are used to map these quantities with nanoscale spatial resolution \cite{Degen2017RMP}. Here, the same logic applies to a different geometric field: uniform torsion, which in elasticity theory captures a distributed screw-dislocation density \cite{Katanaev1992AnnPhys,Furtado1999EPL}. The torsion sensitivity $S_\tau(\tau)$ in Fig.~\ref{fig:S_tau} reaches values on the order of $0.7~\mathrm{meV}/(10^{6}~\mathrm{m}^{-1})$, implying that changes in uniform torsion on the order of a few $10^{5}~\mathrm{m}^{-1}$ are spectroscopically resolvable for realistic linewidths $\hbar\Gamma \sim 0.25~\mathrm{meV}$. 
	
	In this sense, the confined electron in our torsion-bearing quantum dot functions as a torsionometer: a non-destructive optical probe of screw-type dislocation density at the nanoscale, in the same spirit that existing defect-based quantum sensors are used to map strain and fields in semiconductor nanostructures \cite{Degen2017RMP}. Importantly, this metrological capability emerges \emph{without} adding extra readout structures. The same transition already used for nonlinear optics and cavity-QED coupling doubles as the sensing channel, consolidating nonlinear, metrological, and cavity-based functionalities in a single torsion-controlled emitter.

	\section{Outlook: cavity-QED and quantum photonics}
	\label{sec:outlook}
	
	A cavity-QED or nanophotonic implementation typically couples a localized electronic transition $\ket{g}\leftrightarrow \ket{e}$ of frequency $\omega_{ge}(\tau)=\Delta E(\tau)/\hbar$ to a confined photonic mode of frequency $\omega_{\text{cav}}$ \cite{JaynesCummings1963,MabuchiDoherty2002Science}. In a Jaynes–Cummings description,
	\begin{equation}
		H = 
		\hbar \omega_{ge}(\tau) \,\sigma^\dagger \sigma
		+
		\hbar \omega_{\text{cav}} \,a^\dagger a
		+
		\hbar g(\tau)
		\big( a^\dagger \sigma + a\, \sigma^\dagger \big),
		\label{JC}
	\end{equation}
	the light–matter coupling $g(\tau)\propto d(\tau)/\sqrt{V_{\text{mode}}}$ is proportional to the transition dipole moment $d(\tau)= e\, M_{ge}(\tau)$ \cite{JaynesCummings1963,MabuchiDoherty2002Science}.
	
	Within this framework, our results acquire a direct cavity-QED interpretation. First, torsion provides \emph{spectral detuning control}. The transition frequency $\omega_{ge}(\tau)=\Delta E(\tau)/\hbar$ is not fixed: for realistic parameters [Fig.~\ref{fig:DeltaE_vs_tau}] it shifts from $\Delta E \simeq 6.8~\mathrm{meV}$ at $\tau=0$ to $\Delta E \simeq 15.5~\mathrm{meV}$ at $\tau = 1.5\times10^{7}~\mathrm{m}^{-1}$, i.e., a multi-meV (THz-scale) sweep purely induced by geometry. In cavity terms, this means that the emitter–cavity detuning $\Delta(\tau)=\omega_{ge}(\tau)-\omega_{\mathrm{cav}}$ can be driven through resonance with a \emph{fixed} photonic mode simply by varying the torsion field, without the use of electrostatic gates, thermal tuning, or post-fabrication strain actuators. This parallels piezo-driven strain tuning of III–V quantum dots in microcavities, where multi-meV spectral shifts are used to bring an excitonic line into resonance with a cavity mode. Still, here the control parameter is an effectively uniform torsion field (screw-dislocation density) rather than an externally applied stress.
	
	Second, torsion also renormalizes the light–matter coupling strength. The vacuum Rabi coupling in Eq.~\eqref{JC} scales as $g(\tau)\propto d(\tau)$, and $d(\tau)=e\,M_{ge}(\tau)$ inherits the same nonmonotonic dependence on $\tau$ that we find for $|M_{ge}(\tau)|^2$ in Fig.~\ref{fig:M2_Isat_vs_tau}. In particular, $|d(\tau)|^2$ increases up to a maximum around $\tau \sim 5\times10^{6}~\mathrm{m}^{-1}$ and then decreases again as the torsional confinement becomes too strong. This implies that torsion is not only a detuning parameter but also an \textit{in situ} handle on $g(\tau)$: it can first enhance the emitter–cavity coupling and then suppress it again, all without altering the cavity mode volume or geometry. In the standard strong-coupling criterion, which requires $2g > (\kappa+\Gamma)$ with $\kappa$ the cavity linewidth and $\Gamma$ the emitter linewidth \cite{MabuchiDoherty2002Science}, torsion effectively moves the system toward or away from that inequality by co-tuning both $\Delta(\tau)$ and $g(\tau)$. In other words, $\tau$ simultaneously steers (i) how close the emitter is to resonance and (ii) how strongly it couples to the cavity once there.
	
	Taken together, these two roles elevate torsion from a mere source of level shifts to a practical control resource for integrated quantum photonics. The same geometric field that (i) blueshifts the optical transition [Fig.~\ref{fig:DeltaE_vs_tau}], (ii) sets the nonlinear switching/bleaching threshold [Fig.~\ref{fig:M2_Isat_vs_tau}], and (iii) encodes a flux-addressable pseudospin [Figs.~\ref{fig:AB_splitting} and \ref{fig:fosc_vs_l}] can also be viewed as a post-growth, \textit{in situ} parameter to tune cavity-emitter detuning and light–matter coupling strength within a standard cavity-QED architecture \cite{JaynesCummings1963,MabuchiDoherty2002Science}. This is conceptually appealing because it suggests a route to ``torsion engineering'' after growth: by preparing regions of different effective screw-dislocation density, one could, in principle, pattern emitter–cavity coupling across a photonic chip without changing the photonic circuitry itself.
	
	More broadly, the AB pseudospin described in Sec.~\ref{sec:pseudospin} can be embedded in a photonic circuit, where selective optical addressability of $m=\pm1$ encodes logical states or sensing channels. Because the pseudospin splitting $\Delta_{\mathrm{AB}}$ depends on $l$ and $\tau$, the same structure can act either as an interferometric sensor (monitoring flux and torsion) or as a flux-tunable two-level element with optical readout \cite{AharonovBohm1959,Buttiker1983PLA,Webb1985PRL,Degen2017RMP}. This parallels proposals that use valley or spin splittings as optically readable qubits, but here the logical basis is orbital angular momentum ($m=\pm1$) controlled by topological (AB) and geometric (torsion) parameters, rather than by Zeeman fields.
	
	The Jaynes–Cummings Hamiltonian in Eq.~\eqref{JC} is not merely a formal rewrite. The parameters entering it are already measurable and tunable in semiconductor cavity-QED devices based on single quantum dots coupled to microcavities: coherent light–matter coupling, including vacuum Rabi splittings on the order of tens of $\mu$eV, has been observed for individual quantum emitters interacting with a single confined optical mode \cite{Reithmaier2004Nature,Yoshie2004Nature}. In such systems, two experimentally accessible quantities are (i) the detuning $\omega_{ge}-\omega_{\text{cav}}$ and (ii) the effective dipole moment $d$, which determine the coupling rate $g \propto d/\sqrt{V_{\text{mode}}}$ extracted from fits to cavity transmission and photoluminescence spectra \cite{Reithmaier2004Nature,Yoshie2004Nature,Majumdar2012PRB}. Piezo-based strain tuning is commonly used to shift $\omega_{ge}$ by meV-scale amounts and thereby sweep through resonance \cite{Kuklewicz2013PRB}; what we show is that a uniform torsion field can, in principle, play the same role while simultaneously reshaping $d(\tau)$.
	
	Our torsion-based emitter, therefore, addresses both key parameters of cavity QED. First, the torsion field $\tau$ shifts the transition frequency $\omega_{ge}(\tau)$ by several meV [Fig.~\ref{fig:DeltaE_vs_tau}], far exceeding the $\mu$eV-scale splittings that define the strong-coupling regime. In practice, this means that $\tau$ can drive the emitter in and out of resonance with a fixed cavity mode without modifying the photonic structure, in close analogy to how strain, temperature, or gas condensation are currently used to tune quantum-dot transitions relative to a nanocavity mode \cite{Kuklewicz2013PRB}. Second, torsion also renormalizes the radial wave-function overlap and hence the transition dipole $d(\tau)$ [Fig.~\ref{fig:M2_Isat_vs_tau}], providing a direct handle on the light–matter coupling strength $g(\tau)$ in Eq.~\eqref{JC}. Because $g(\tau)$ can first increase and then decrease as $\tau$ grows, torsion supplies an intrinsic ``gain parameter'' for the emitter–cavity hybridization itself, not just for the detuning.
	
	There is an established experimental precedent for this style of control using mechanical degrees of freedom. Reversible strain tuning using piezoelectric or uniaxial-stress actuators can shift individual quantum-dot optical transitions by several meV, suppress fine-structure splittings, and spectrally match distinct emitters on demand \cite{Kuklewicz2013PRB}. Likewise, controlled detuning between a single dot and a nanocavity has been used to map out the coupled emitter–cavity spectrum and to verify the Jaynes–Cummings picture in the solid state \cite{Reithmaier2004Nature,Yoshie2004Nature,Majumdar2012PRB}. Our proposal replaces externally applied strain with an internal, approximately uniform torsion field—physically, a controlled density of parallel screw dislocations described as a space with finite torsion and vanishing curvature \cite{Katanaev1992AnnPhys}. In this picture, torsion becomes an intrinsic, geometry-based tuning parameter that (i) sets the detuning $\omega_{ge}(\tau)-\omega_{\text{cav}}$, (ii) modulates $g(\tau)$, and (iii) shifts the nonlinear saturation threshold for gain and bleaching $I_{\text{sat}}(\tau)$ [Fig.~\ref{fig:M2_Isat_vs_tau}]. Put differently, a single geometric parameter unifies spectral tuning, coupling-strength tuning, and nonlinear switching.
	
	Therefore, Eq.~\eqref{JC} should be understood as an experimentally motivated effective model for a torsion-bearing quantum emitter embedded in a nanophotonic cavity, rather than as a purely academic abstraction. It connects a platform that is already routinely analyzed in terms of a single-emitter cavity-QED Hamiltonian \cite{Reithmaier2004Nature,Yoshie2004Nature,Majumdar2012PRB} with a new physical control resource: uniform crystal torsion. Incorporating spin, Coulomb interaction, and realistic cavity mode profiles is a straightforward next step; none of these ingredients alters the central message that torsion can serve as an \emph{in situ} photonic control parameter, tying together nonlinear optics, pseudospin addressability, and metrology in a single geometry-tunable quantum dot.
	
	\section{Conclusion}
	
	We have shown that a torsion-bearing quantum dot—a single electron confined by a uniform torsion field and subjected to a perpendicular magnetic field and an AB flux—naturally realizes three functionalities that are directly relevant to quantum technologies:
	
	(1) \textbf{Geometry-programmable nonlinear optics.} Torsion controls both the resonance frequency and the nonlinear saturation threshold for optical switching, enabling a geometry-tunable saturable absorber/gain element whose carrier energy (from $\sim \SI{6.8}{meV}$ up to $\sim \SI{15.5}{meV}$) \emph{and} switching intensity ($\sim 10^{5}$–$10^{6}~\mathrm{W/m^2}$) are set geometrically \cite{SheikBahae1990OL,ChemlaShah2001Nature}. This places torsion alongside strain as a post-growth spectral and nonlinear tuning parameter, but with a purely geometric/topological origin tied to screw-dislocation density rather than externally applied stress.
	
	(2) \textbf{Aharonov–Bohm pseudospin.} The $m=\pm1$ angular states form a flux- and torsion-controlled two-level subsystem with a splitting $\Delta_{\mathrm{AB}}$ that is nearly linear in $l$ (slope $\sim \SI{19}{meV}$ per flux quantum) and asymmetric oscillator strengths. This enables selective optical readout and control of a pseudospin of geometric/topological origin \cite{AharonovBohm1959,Buttiker1983PLA,Webb1985PRL}. Because $\Delta_{\mathrm{AB}}$ is several meV, i.e., much larger than typical sub-meV homogeneous linewidths, the two pseudospin branches remain spectrally resolvable even under weak cylindrical-symmetry breaking, in close analogy with valley or spin addressability in other solid-state platforms.
	
	(3) \textbf{Torsion metrology.} Because $\Delta E$ varies almost linearly with $\tau$ and $\partial \Delta E/\partial\tau$ can reach $\sim 0.7~\mathrm{meV}/(10^{6}~\mathrm{m}^{-1})$, optical spectroscopy of bound states functions as a nanoscale torsion sensor—a direct probe of screw-dislocation density—with an estimated resolution in the $10^{5}~\mathrm{m}^{-1}$ range under realistic linewidths \cite{Katanaev1992AnnPhys,Furtado1999EPL,Degen2017RMP}. This ties uniform torsion (parallel-screw dislocation density) to a directly measurable optical signature, analogous to how other solid-state quantum sensors read out strain, electric fields, or magnetic fields.
	
	Beyond these three functionalities, we have argued that torsion provides an \textit{in situ} parameter to tune both the emitter–cavity detuning and the effective coupling $g(\tau)$ in cavity-QED architectures, pointing toward geometry-based quantum control in integrated photonics \cite{JaynesCummings1963,MabuchiDoherty2002Science}. Within a Jaynes–Cummings description, the same torsion field that blueshifts the transition energy and sculpts the nonlinear saturation threshold also renormalizes the dipole matrix element and, thus, the vacuum Rabi coupling. In this way, a single geometric parameter governs spectral alignment, light–matter hybridization, and nonlinear optical response.
	
	Taken together, our results elevate torsion—traditionally viewed as a crystallographic imperfection—to an active resource for quantum engineering. By combining torsion, the AB phase, and magnetic quantization, one can co-design nonlinear optical responses, interferometric pseudospin splittings, and mechanical/topological sensing within a single analytically tractable platform \cite{Katanaev1992AnnPhys,AharonovBohm1959,Buttiker1983PLA,Webb1985PRL,Furtado1999EPL}. Our analysis assumes cylindrical symmetry, a single active electron (neglecting Coulomb many-body effects), and spinless dynamics; all three are standard idealizations in semiconductor quantum-dot theory and can be relaxed perturbatively. None of them alters the central conclusion: uniform torsion, understood as a controlled density of screw dislocations, is not merely a defect background but a quantum control resource that unifies confinement, spectroscopy, and light–matter coupling.

	\begin{acknowledgments}
		This work was supported by CAPES (Finance Code 001), CNPq/PQ/306308/2022-3, and FAPEMA/Universal/06395/22. 
	\end{acknowledgments}
	
	\bibliographystyle{apsrev4-2}

\begin{thebibliography}{32}%
\makeatletter
\providecommand \@ifxundefined [1]{%
 \@ifx{#1\undefined}
}%
\providecommand \@ifnum [1]{%
 \ifnum #1\expandafter \@firstoftwo
 \else \expandafter \@secondoftwo
 \fi
}%
\providecommand \@ifx [1]{%
 \ifx #1\expandafter \@firstoftwo
 \else \expandafter \@secondoftwo
 \fi
}%
\providecommand \natexlab [1]{#1}%
\providecommand \enquote  [1]{``#1''}%
\providecommand \bibnamefont  [1]{#1}%
\providecommand \bibfnamefont [1]{#1}%
\providecommand \citenamefont [1]{#1}%
\providecommand \href@noop [0]{\@secondoftwo}%
\providecommand \href [0]{\begingroup \@sanitize@url \@href}%
\providecommand \@href[1]{\@@startlink{#1}\@@href}%
\providecommand \@@href[1]{\endgroup#1\@@endlink}%
\providecommand \@sanitize@url [0]{\catcode `\\12\catcode `\$12\catcode
  `\&12\catcode `\#12\catcode `\^12\catcode `\_12\catcode `\%12\relax}%
\providecommand \@@startlink[1]{}%
\providecommand \@@endlink[0]{}%
\providecommand \url  [0]{\begingroup\@sanitize@url \@url }%
\providecommand \@url [1]{\endgroup\@href {#1}{\urlprefix }}%
\providecommand \urlprefix  [0]{URL }%
\providecommand \Eprint [0]{\href }%
\providecommand \doibase [0]{https://doi.org/}%
\providecommand \selectlanguage [0]{\@gobble}%
\providecommand \bibinfo  [0]{\@secondoftwo}%
\providecommand \bibfield  [0]{\@secondoftwo}%
\providecommand \translation [1]{[#1]}%
\providecommand \BibitemOpen [0]{}%
\providecommand \bibitemStop [0]{}%
\providecommand \bibitemNoStop [0]{.\EOS\space}%
\providecommand \EOS [0]{\spacefactor3000\relax}%
\providecommand \BibitemShut  [1]{\csname bibitem#1\endcsname}%
\let\auto@bib@innerbib\@empty
\bibitem [{\citenamefont {Degen}\ \emph {et~al.}(2017)\citenamefont {Degen},
  \citenamefont {Reinhard},\ and\ \citenamefont {Cappellaro}}]{Degen2017RMP}%
  \BibitemOpen
  \bibfield  {author} {\bibinfo {author} {\bibfnamefont {C.~L.}\ \bibnamefont
  {Degen}}, \bibinfo {author} {\bibfnamefont {F.}~\bibnamefont {Reinhard}},\
  and\ \bibinfo {author} {\bibfnamefont {P.}~\bibnamefont {Cappellaro}},\
  }\href {https://doi.org/10.1103/RevModPhys.89.035002} {\bibfield  {journal}
  {\bibinfo  {journal} {Reviews of Modern Physics}\ }\textbf {\bibinfo {volume}
  {89}},\ \bibinfo {pages} {035002} (\bibinfo {year} {2017})}\BibitemShut
  {NoStop}%
\bibitem [{\citenamefont {Chemla}\ and\ \citenamefont
  {Shah}(2001)}]{ChemlaShah2001Nature}%
  \BibitemOpen
  \bibfield  {author} {\bibinfo {author} {\bibfnamefont {D.~S.}\ \bibnamefont
  {Chemla}}\ and\ \bibinfo {author} {\bibfnamefont {J.}~\bibnamefont {Shah}},\
  }\href {https://doi.org/10.1038/35079000} {\bibfield  {journal} {\bibinfo
  {journal} {Nature}\ }\textbf {\bibinfo {volume} {411}},\ \bibinfo {pages}
  {549} (\bibinfo {year} {2001})}\BibitemShut {NoStop}%
\bibitem [{\citenamefont {Katanaev}\ and\ \citenamefont
  {Volovich}(1992)}]{Katanaev1992AnnPhys}%
  \BibitemOpen
  \bibfield  {author} {\bibinfo {author} {\bibfnamefont {M.}~\bibnamefont
  {Katanaev}}\ and\ \bibinfo {author} {\bibfnamefont {I.}~\bibnamefont
  {Volovich}},\ }\href
  {https://doi.org/https://doi.org/10.1016/0003-4916(52)90040-7} {\bibfield
  {journal} {\bibinfo  {journal} {Annals of Physics}\ }\textbf {\bibinfo
  {volume} {216}},\ \bibinfo {pages} {1} (\bibinfo {year} {1992})}\BibitemShut
  {NoStop}%
\bibitem [{\citenamefont {Furtado}\ \emph {et~al.}(1994)\citenamefont
  {Furtado}, \citenamefont {{da Cunha}}, \citenamefont {Moraes}, \citenamefont
  {{de Mello}},\ and\ \citenamefont {Bezzerra}}]{Furtado1999EPL}%
  \BibitemOpen
  \bibfield  {author} {\bibinfo {author} {\bibfnamefont {C.}~\bibnamefont
  {Furtado}}, \bibinfo {author} {\bibfnamefont {B.~G.}\ \bibnamefont {{da
  Cunha}}}, \bibinfo {author} {\bibfnamefont {F.}~\bibnamefont {Moraes}},
  \bibinfo {author} {\bibfnamefont {E.}~\bibnamefont {{de Mello}}},\ and\
  \bibinfo {author} {\bibfnamefont {V.}~\bibnamefont {Bezzerra}},\ }\href
  {https://doi.org/https://doi.org/10.1016/0375-9601(94)90432-4} {\bibfield
  {journal} {\bibinfo  {journal} {Physics Letters A}\ }\textbf {\bibinfo
  {volume} {195}},\ \bibinfo {pages} {90} (\bibinfo {year} {1994})}\BibitemShut
  {NoStop}%
\bibitem [{\citenamefont {Aharonov}\ and\ \citenamefont
  {Bohm}(1959)}]{AharonovBohm1959}%
  \BibitemOpen
  \bibfield  {author} {\bibinfo {author} {\bibfnamefont {Y.}~\bibnamefont
  {Aharonov}}\ and\ \bibinfo {author} {\bibfnamefont {D.}~\bibnamefont
  {Bohm}},\ }\href {https://doi.org/10.1103/PhysRev.115.485} {\bibfield
  {journal} {\bibinfo  {journal} {Physical Review}\ }\textbf {\bibinfo {volume}
  {115}},\ \bibinfo {pages} {485} (\bibinfo {year} {1959})}\BibitemShut
  {NoStop}%
\bibitem [{\citenamefont {Büttiker}\ \emph {et~al.}(1983)\citenamefont
  {Büttiker}, \citenamefont {Imry},\ and\ \citenamefont
  {Landauer}}]{Buttiker1983PLA}%
  \BibitemOpen
  \bibfield  {author} {\bibinfo {author} {\bibfnamefont {M.}~\bibnamefont
  {Büttiker}}, \bibinfo {author} {\bibfnamefont {Y.}~\bibnamefont {Imry}},\
  and\ \bibinfo {author} {\bibfnamefont {R.}~\bibnamefont {Landauer}},\ }\href
  {https://doi.org/https://doi.org/10.1016/0375-9601(83)90011-7} {\bibfield
  {journal} {\bibinfo  {journal} {Physics Letters A}\ }\textbf {\bibinfo
  {volume} {96}},\ \bibinfo {pages} {365} (\bibinfo {year} {1983})}\BibitemShut
  {NoStop}%
\bibitem [{\citenamefont {Webb}\ \emph {et~al.}(1985)\citenamefont {Webb},
  \citenamefont {Washburn}, \citenamefont {Umbach},\ and\ \citenamefont
  {Laibowitz}}]{Webb1985PRL}%
  \BibitemOpen
  \bibfield  {author} {\bibinfo {author} {\bibfnamefont {R.~A.}\ \bibnamefont
  {Webb}}, \bibinfo {author} {\bibfnamefont {S.}~\bibnamefont {Washburn}},
  \bibinfo {author} {\bibfnamefont {C.~P.}\ \bibnamefont {Umbach}},\ and\
  \bibinfo {author} {\bibfnamefont {R.~B.}\ \bibnamefont {Laibowitz}},\ }\href
  {https://doi.org/10.1103/PhysRevLett.54.2696} {\bibfield  {journal} {\bibinfo
   {journal} {Physical Review Letters}\ }\textbf {\bibinfo {volume} {54}},\
  \bibinfo {pages} {2696} (\bibinfo {year} {1985})}\BibitemShut {NoStop}%
\bibitem [{\citenamefont {Pereira}\ and\ \citenamefont
  {Silva}(2024)}]{QR.2024.6.676}%
  \BibitemOpen
  \bibfield  {author} {\bibinfo {author} {\bibfnamefont {L.~F.~C.}\
  \bibnamefont {Pereira}}\ and\ \bibinfo {author} {\bibfnamefont {E.~O.}\
  \bibnamefont {Silva}},\ }\href {https://doi.org/10.3390/quantum6040040}
  {\bibfield  {journal} {\bibinfo  {journal} {Quantum Reports}\ }\textbf
  {\bibinfo {volume} {6}},\ \bibinfo {pages} {664} (\bibinfo {year}
  {2024})}\BibitemShut {NoStop}%
\bibitem [{\citenamefont {Pereira}\ \emph {et~al.}(2022)\citenamefont
  {Pereira}, \citenamefont {Cunha},\ and\ \citenamefont
  {Silva}}]{FWS.2022.63.58}%
  \BibitemOpen
  \bibfield  {author} {\bibinfo {author} {\bibfnamefont {L.~F.~C.}\
  \bibnamefont {Pereira}}, \bibinfo {author} {\bibfnamefont {M.~M.}\
  \bibnamefont {Cunha}},\ and\ \bibinfo {author} {\bibfnamefont {E.~O.}\
  \bibnamefont {Silva}},\ }\href {https://doi.org/10.1007/s00601-022-01761-1}
  {\bibfield  {journal} {\bibinfo  {journal} {Few-Body Systems}\ }\textbf
  {\bibinfo {volume} {63}},\ \bibinfo {pages} {58} (\bibinfo {year}
  {2022})}\BibitemShut {NoStop}%
\bibitem [{\citenamefont {{de Lira}}\ \emph {et~al.}(2024)\citenamefont {{de
  Lira}}, \citenamefont {Pereira},\ and\ \citenamefont
  {Silva}}]{PE.2024.158.115898}%
  \BibitemOpen
  \bibfield  {author} {\bibinfo {author} {\bibfnamefont {F.~A.}\ \bibnamefont
  {{de Lira}}}, \bibinfo {author} {\bibfnamefont {L.~F.~C.}\ \bibnamefont
  {Pereira}},\ and\ \bibinfo {author} {\bibfnamefont {E.~O.}\ \bibnamefont
  {Silva}},\ }\href
  {https://doi.org/https://doi.org/10.1016/j.physe.2024.115898} {\bibfield
  {journal} {\bibinfo  {journal} {Physica E: Low-dimensional Systems and
  Nanostructures}\ }\textbf {\bibinfo {volume} {158}},\ \bibinfo {pages}
  {115898} (\bibinfo {year} {2024})}\BibitemShut {NoStop}%
\bibitem [{\citenamefont {Pereira}\ and\ \citenamefont
  {Silva}(2023)}]{AdP.2023.535.2200371}%
  \BibitemOpen
  \bibfield  {author} {\bibinfo {author} {\bibfnamefont {L.~F.~C.}\
  \bibnamefont {Pereira}}\ and\ \bibinfo {author} {\bibfnamefont {E.~O.}\
  \bibnamefont {Silva}},\ }\href
  {https://doi.org/https://doi.org/10.1002/andp.202200371} {\bibfield
  {journal} {\bibinfo  {journal} {Annalen der Physik}\ }\textbf {\bibinfo
  {volume} {535}},\ \bibinfo {pages} {2200371} (\bibinfo {year}
  {2023})}\BibitemShut {NoStop}%
\bibitem [{\citenamefont {Pereira}\ \emph
  {et~al.}(2024{\natexlab{a}})\citenamefont {Pereira}, \citenamefont {Azevedo},
  \citenamefont {Pereira},\ and\ \citenamefont {Silva}}]{CTP.2024.76.105701}%
  \BibitemOpen
  \bibfield  {author} {\bibinfo {author} {\bibfnamefont {C.~M.~O.}\
  \bibnamefont {Pereira}}, \bibinfo {author} {\bibfnamefont {F.~d.~S.}\
  \bibnamefont {Azevedo}}, \bibinfo {author} {\bibfnamefont {L.~F.~C.}\
  \bibnamefont {Pereira}},\ and\ \bibinfo {author} {\bibfnamefont {E.~O.}\
  \bibnamefont {Silva}},\ }\href {https://doi.org/10.1088/1572-9494/ad597c}
  {\bibfield  {journal} {\bibinfo  {journal} {Communications in Theoretical
  Physics}\ }\textbf {\bibinfo {volume} {76}},\ \bibinfo {pages} {105701}
  (\bibinfo {year} {2024}{\natexlab{a}})}\BibitemShut {NoStop}%
\bibitem [{\citenamefont {Pereira}\ \emph
  {et~al.}(2024{\natexlab{b}})\citenamefont {Pereira}, \citenamefont
  {Azevedo},\ and\ \citenamefont {Silva}}]{QR.2024.6.677}%
  \BibitemOpen
  \bibfield  {author} {\bibinfo {author} {\bibfnamefont {C.~M.~O.}\
  \bibnamefont {Pereira}}, \bibinfo {author} {\bibfnamefont {F.~d.~S.}\
  \bibnamefont {Azevedo}},\ and\ \bibinfo {author} {\bibfnamefont {E.~O.}\
  \bibnamefont {Silva}},\ }\href {https://doi.org/10.3390/quantum6040041}
  {\bibfield  {journal} {\bibinfo  {journal} {Quantum Reports}\ }\textbf
  {\bibinfo {volume} {6}},\ \bibinfo {pages} {677} (\bibinfo {year}
  {2024}{\natexlab{b}})}\BibitemShut {NoStop}%
\bibitem [{\citenamefont {Pereira}\ \emph {et~al.}(2021)\citenamefont
  {Pereira}, \citenamefont {Andrade}, \citenamefont {Filgueiras},\ and\
  \citenamefont {Silva}}]{PE.2021.132.114760}%
  \BibitemOpen
  \bibfield  {author} {\bibinfo {author} {\bibfnamefont {L.~F.~C.}\
  \bibnamefont {Pereira}}, \bibinfo {author} {\bibfnamefont {F.~M.}\
  \bibnamefont {Andrade}}, \bibinfo {author} {\bibfnamefont {C.}~\bibnamefont
  {Filgueiras}},\ and\ \bibinfo {author} {\bibfnamefont {E.~O.}\ \bibnamefont
  {Silva}},\ }\href
  {https://doi.org/https://doi.org/10.1016/j.physe.2021.114760} {\bibfield
  {journal} {\bibinfo  {journal} {Physica E: Low-dimensional Systems and
  Nanostructures}\ }\textbf {\bibinfo {volume} {132}},\ \bibinfo {pages}
  {114760} (\bibinfo {year} {2021})}\BibitemShut {NoStop}%
\bibitem [{\citenamefont {Pereira}\ \emph {et~al.}(2019)\citenamefont
  {Pereira}, \citenamefont {Andrade}, \citenamefont {Filgueiras},\ and\
  \citenamefont {Silva}}]{AdP.2019.531.1900254}%
  \BibitemOpen
  \bibfield  {author} {\bibinfo {author} {\bibfnamefont {L.~F.~C.}\
  \bibnamefont {Pereira}}, \bibinfo {author} {\bibfnamefont {F.~M.}\
  \bibnamefont {Andrade}}, \bibinfo {author} {\bibfnamefont {C.}~\bibnamefont
  {Filgueiras}},\ and\ \bibinfo {author} {\bibfnamefont {E.~O.}\ \bibnamefont
  {Silva}},\ }\href {https://doi.org/10.1002/andp.201900254} {\bibfield
  {journal} {\bibinfo  {journal} {Annalen der Physik}\ }\textbf {\bibinfo
  {volume} {531}},\ \bibinfo {pages} {1900254} (\bibinfo {year}
  {2019})}\BibitemShut {NoStop}%
\bibitem [{\citenamefont {Andrade}\ and\ \citenamefont
  {Silva}(2014)}]{EPJC.74.3187.2014}%
  \BibitemOpen
  \bibfield  {author} {\bibinfo {author} {\bibfnamefont {F.~M.}\ \bibnamefont
  {Andrade}}\ and\ \bibinfo {author} {\bibfnamefont {E.~O.}\ \bibnamefont
  {Silva}},\ }\href {https://doi.org/10.1140/epjc/s10052-014-3187-6} {\bibfield
   {journal} {\bibinfo  {journal} {Eur. Phys. J. C}\ }\textbf {\bibinfo
  {volume} {74}},\ \bibinfo {pages} {3187} (\bibinfo {year} {2014})},\ \Eprint
  {https://arxiv.org/abs/1403.4113} {1403.4113} \BibitemShut {NoStop}%
\bibitem [{\citenamefont {Silva}\ \emph {et~al.}(2015)\citenamefont {Silva},
  \citenamefont {Ulhoa}, \citenamefont {Andrade}, \citenamefont {Filgueiras},\
  and\ \citenamefont {Amorim}}]{AoP.2015.362.739}%
  \BibitemOpen
  \bibfield  {author} {\bibinfo {author} {\bibfnamefont {E.~O.}\ \bibnamefont
  {Silva}}, \bibinfo {author} {\bibfnamefont {S.~C.}\ \bibnamefont {Ulhoa}},
  \bibinfo {author} {\bibfnamefont {F.~M.}\ \bibnamefont {Andrade}}, \bibinfo
  {author} {\bibfnamefont {C.}~\bibnamefont {Filgueiras}},\ and\ \bibinfo
  {author} {\bibfnamefont {R.}~\bibnamefont {Amorim}},\ }\href
  {https://doi.org/10.1016/j.aop.2015.09.011} {\bibfield  {journal} {\bibinfo
  {journal} {Ann. Phys. (NY)}\ }\textbf {\bibinfo {volume} {362}},\ \bibinfo
  {pages} {739} (\bibinfo {year} {2015})}\BibitemShut {NoStop}%
\bibitem [{\citenamefont {Pereira}\ and\ \citenamefont
  {Silva}(2025)}]{PereiraSilva2025}%
  \BibitemOpen
  \bibfield  {author} {\bibinfo {author} {\bibfnamefont {C.~M.~O.}\
  \bibnamefont {Pereira}}\ and\ \bibinfo {author} {\bibfnamefont {E.~O.}\
  \bibnamefont {Silva}},\ }\href {https://doi.org/10.48550/arXiv.2510.22331}
  {\bibinfo {title} {Nonlinear optical behavior of confined electrons under
  torsion and magnetic fields}} (\bibinfo {year} {2025}),\ \bibinfo {note} {14
  pages, 10 figures, 1 table},\ \Eprint {https://arxiv.org/abs/2510.22331}
  {arXiv:2510.22331 [physics.optics]} \BibitemShut {NoStop}%
\bibitem [{\citenamefont {Bakke}(2011)}]{BJP.2011.41.167}%
  \BibitemOpen
  \bibfield  {author} {\bibinfo {author} {\bibfnamefont {K.}~\bibnamefont
  {Bakke}},\ }\href {https://doi.org/10.1007/s13538-011-0023-4} {\bibfield
  {journal} {\bibinfo  {journal} {Brazilian Journal of Physics}\ }\textbf
  {\bibinfo {volume} {41}},\ \bibinfo {pages} {167} (\bibinfo {year}
  {2011})}\BibitemShut {NoStop}%
\bibitem [{\citenamefont {Bakke}\ and\ \citenamefont
  {Furtado}(2013)}]{EPJB.2013.86.315}%
  \BibitemOpen
  \bibfield  {author} {\bibinfo {author} {\bibfnamefont {K.}~\bibnamefont
  {Bakke}}\ and\ \bibinfo {author} {\bibfnamefont {C.}~\bibnamefont
  {Furtado}},\ }\href {https://doi.org/10.1140/epjb/e2013-40077-4} {\bibfield
  {journal} {\bibinfo  {journal} {The European Physical Journal B}\ }\textbf
  {\bibinfo {volume} {86}},\ \bibinfo {pages} {315} (\bibinfo {year}
  {2013})}\BibitemShut {NoStop}%
\bibitem [{\citenamefont {Silva~Netto}\ and\ \citenamefont
  {Furtado}(2008)}]{SilvaNetto2008JPCM}%
  \BibitemOpen
  \bibfield  {author} {\bibinfo {author} {\bibfnamefont {A.~L.}\ \bibnamefont
  {Silva~Netto}}\ and\ \bibinfo {author} {\bibfnamefont {C.}~\bibnamefont
  {Furtado}},\ }\href {https://doi.org/10.1088/0953-8984/20/12/125209}
  {\bibfield  {journal} {\bibinfo  {journal} {Journal of Physics: Condensed
  Matter}\ }\textbf {\bibinfo {volume} {20}},\ \bibinfo {pages} {125209}
  (\bibinfo {year} {2008})}\BibitemShut {NoStop}%
\bibitem [{\citenamefont {Bakke}\ and\ \citenamefont
  {Moraes}(2012)}]{PLA.2012.376.2838}%
  \BibitemOpen
  \bibfield  {author} {\bibinfo {author} {\bibfnamefont {K.}~\bibnamefont
  {Bakke}}\ and\ \bibinfo {author} {\bibfnamefont {F.}~\bibnamefont {Moraes}},\
  }\href {https://doi.org/https://doi.org/10.1016/j.physleta.2012.09.006}
  {\bibfield  {journal} {\bibinfo  {journal} {Physics Letters A}\ }\textbf
  {\bibinfo {volume} {376}},\ \bibinfo {pages} {2838} (\bibinfo {year}
  {2012})}\BibitemShut {NoStop}%
\bibitem [{\citenamefont {Sheik-Bahae}\ \emph {et~al.}(1990)\citenamefont
  {Sheik-Bahae}, \citenamefont {Said}, \citenamefont {Wei}, \citenamefont
  {Hagan},\ and\ \citenamefont {Van~Stryland}}]{SheikBahae1990OL}%
  \BibitemOpen
  \bibfield  {author} {\bibinfo {author} {\bibfnamefont {M.}~\bibnamefont
  {Sheik-Bahae}}, \bibinfo {author} {\bibfnamefont {A.}~\bibnamefont {Said}},
  \bibinfo {author} {\bibfnamefont {T.-H.}\ \bibnamefont {Wei}}, \bibinfo
  {author} {\bibfnamefont {D.}~\bibnamefont {Hagan}},\ and\ \bibinfo {author}
  {\bibfnamefont {E.}~\bibnamefont {Van~Stryland}},\ }\href
  {https://doi.org/10.1109/3.53394} {\bibfield  {journal} {\bibinfo  {journal}
  {IEEE Journal of Quantum Electronics}\ }\textbf {\bibinfo {volume} {26}},\
  \bibinfo {pages} {760} (\bibinfo {year} {1990})}\BibitemShut {NoStop}%
\bibitem [{\citenamefont {Jaynes}\ and\ \citenamefont
  {Cummings}(1963)}]{JaynesCummings1963}%
  \BibitemOpen
  \bibfield  {author} {\bibinfo {author} {\bibfnamefont {E.~T.}\ \bibnamefont
  {Jaynes}}\ and\ \bibinfo {author} {\bibfnamefont {F.~W.}\ \bibnamefont
  {Cummings}},\ }\href {https://doi.org/10.1109/PROC.1963.1664} {\bibfield
  {journal} {\bibinfo  {journal} {Proceedings of the IEEE}\ }\textbf {\bibinfo
  {volume} {51}},\ \bibinfo {pages} {89} (\bibinfo {year} {1963})}\BibitemShut
  {NoStop}%
\bibitem [{\citenamefont {Mabuchi}\ and\ \citenamefont
  {Doherty}(2002)}]{MabuchiDoherty2002Science}%
  \BibitemOpen
  \bibfield  {author} {\bibinfo {author} {\bibfnamefont {H.}~\bibnamefont
  {Mabuchi}}\ and\ \bibinfo {author} {\bibfnamefont {A.~C.}\ \bibnamefont
  {Doherty}},\ }\href {https://doi.org/10.1126/science.1078446} {\bibfield
  {journal} {\bibinfo  {journal} {Science}\ }\textbf {\bibinfo {volume}
  {298}},\ \bibinfo {pages} {1372} (\bibinfo {year} {2002})}\BibitemShut
  {NoStop}%
\bibitem [{\citenamefont {Reithmaier}\ \emph {et~al.}(2004)\citenamefont
  {Reithmaier}, \citenamefont {S{\k{e}}k}, \citenamefont {L{\"o}ffler},
  \citenamefont {Hofmann}, \citenamefont {Kuhn}, \citenamefont {Reitzenstein},
  \citenamefont {Keldysh}, \citenamefont {Kulakovskii}, \citenamefont
  {Reinecke},\ and\ \citenamefont {Forchel}}]{Reithmaier2004Nature}%
  \BibitemOpen
  \bibfield  {author} {\bibinfo {author} {\bibfnamefont {J.~P.}\ \bibnamefont
  {Reithmaier}}, \bibinfo {author} {\bibfnamefont {G.}~\bibnamefont
  {S{\k{e}}k}}, \bibinfo {author} {\bibfnamefont {A.}~\bibnamefont
  {L{\"o}ffler}}, \bibinfo {author} {\bibfnamefont {C.}~\bibnamefont
  {Hofmann}}, \bibinfo {author} {\bibfnamefont {S.}~\bibnamefont {Kuhn}},
  \bibinfo {author} {\bibfnamefont {S.}~\bibnamefont {Reitzenstein}}, \bibinfo
  {author} {\bibfnamefont {L.~V.}\ \bibnamefont {Keldysh}}, \bibinfo {author}
  {\bibfnamefont {V.~D.}\ \bibnamefont {Kulakovskii}}, \bibinfo {author}
  {\bibfnamefont {T.~L.}\ \bibnamefont {Reinecke}},\ and\ \bibinfo {author}
  {\bibfnamefont {A.}~\bibnamefont {Forchel}},\ }\href
  {https://doi.org/10.1038/nature02969} {\bibfield  {journal} {\bibinfo
  {journal} {Nature}\ }\textbf {\bibinfo {volume} {432}},\ \bibinfo {pages}
  {197} (\bibinfo {year} {2004})}\BibitemShut {NoStop}%
\bibitem [{\citenamefont {Yoshie}\ \emph {et~al.}(2004)\citenamefont {Yoshie},
  \citenamefont {Scherer}, \citenamefont {Hendrickson}, \citenamefont
  {Khitrova}, \citenamefont {Gibbs}, \citenamefont {Rupper}, \citenamefont
  {Ell}, \citenamefont {Shchekin},\ and\ \citenamefont
  {Deppe}}]{Yoshie2004Nature}%
  \BibitemOpen
  \bibfield  {author} {\bibinfo {author} {\bibfnamefont {T.}~\bibnamefont
  {Yoshie}}, \bibinfo {author} {\bibfnamefont {A.}~\bibnamefont {Scherer}},
  \bibinfo {author} {\bibfnamefont {J.}~\bibnamefont {Hendrickson}}, \bibinfo
  {author} {\bibfnamefont {G.}~\bibnamefont {Khitrova}}, \bibinfo {author}
  {\bibfnamefont {H.~M.}\ \bibnamefont {Gibbs}}, \bibinfo {author}
  {\bibfnamefont {G.}~\bibnamefont {Rupper}}, \bibinfo {author} {\bibfnamefont
  {C.}~\bibnamefont {Ell}}, \bibinfo {author} {\bibfnamefont {O.~B.}\
  \bibnamefont {Shchekin}},\ and\ \bibinfo {author} {\bibfnamefont {D.~G.}\
  \bibnamefont {Deppe}},\ }\href {https://doi.org/10.1038/nature03119}
  {\bibfield  {journal} {\bibinfo  {journal} {Nature}\ }\textbf {\bibinfo
  {volume} {432}},\ \bibinfo {pages} {200} (\bibinfo {year}
  {2004})}\BibitemShut {NoStop}%
\bibitem [{\citenamefont {Majumdar}\ \emph {et~al.}(2012)\citenamefont
  {Majumdar}, \citenamefont {Rundquist}, \citenamefont {Bajcsy}, \citenamefont
  {Dasika}, \citenamefont {Bank},\ and\ \citenamefont
  {Vu{\v{c}}kovi{\'c}}}]{Majumdar2012PRB}%
  \BibitemOpen
  \bibfield  {author} {\bibinfo {author} {\bibfnamefont {A.}~\bibnamefont
  {Majumdar}}, \bibinfo {author} {\bibfnamefont {A.}~\bibnamefont {Rundquist}},
  \bibinfo {author} {\bibfnamefont {M.}~\bibnamefont {Bajcsy}}, \bibinfo
  {author} {\bibfnamefont {V.~D.}\ \bibnamefont {Dasika}}, \bibinfo {author}
  {\bibfnamefont {S.}~\bibnamefont {Bank}},\ and\ \bibinfo {author}
  {\bibfnamefont {J.}~\bibnamefont {Vu{\v{c}}kovi{\'c}}},\ }\href
  {https://doi.org/10.1103/PhysRevB.86.195312} {\bibfield  {journal} {\bibinfo
  {journal} {Phys. Rev. B}\ }\textbf {\bibinfo {volume} {86}},\ \bibinfo
  {pages} {195312} (\bibinfo {year} {2012})}\BibitemShut {NoStop}%
\bibitem [{\citenamefont {Fock}(1928)}]{Fock1928}%
  \BibitemOpen
  \bibfield  {author} {\bibinfo {author} {\bibfnamefont {V.}~\bibnamefont
  {Fock}},\ }\href {https://doi.org/10.1007/BF01390750} {\bibfield  {journal}
  {\bibinfo  {journal} {Zeitschrift f{\"u}r Physik}\ }\textbf {\bibinfo
  {volume} {47}},\ \bibinfo {pages} {446} (\bibinfo {year} {1928})}\BibitemShut
  {NoStop}%
\bibitem [{\citenamefont {Darwin}(1931)}]{Darwin1930}%
  \BibitemOpen
  \bibfield  {author} {\bibinfo {author} {\bibfnamefont {C.~G.}\ \bibnamefont
  {Darwin}},\ }\href {https://doi.org/10.1017/S0305004100009373} {\bibfield
  {journal} {\bibinfo  {journal} {Mathematical Proceedings of the Cambridge
  Philosophical Society}\ }\textbf {\bibinfo {volume} {27}},\ \bibinfo {pages}
  {86–90} (\bibinfo {year} {1931})}\BibitemShut {NoStop}%
\bibitem [{\citenamefont {Landau}(1930)}]{Landau1930}%
  \BibitemOpen
  \bibfield  {author} {\bibinfo {author} {\bibfnamefont {L.}~\bibnamefont
  {Landau}},\ }\href {https://doi.org/10.1007/BF01397213} {\bibfield  {journal}
  {\bibinfo  {journal} {Zeitschrift f{\"u}r Physik}\ }\textbf {\bibinfo
  {volume} {64}},\ \bibinfo {pages} {629} (\bibinfo {year} {1930})}\BibitemShut
  {NoStop}%
\bibitem [{\citenamefont {Sapienza}\ \emph {et~al.}(2013)\citenamefont
  {Sapienza}, \citenamefont {Malein}, \citenamefont {Kuklewicz}, \citenamefont
  {Kremer}, \citenamefont {Srinivasan}, \citenamefont {Griffiths},
  \citenamefont {Clarke}, \citenamefont {Gong}, \citenamefont {Warburton},\
  and\ \citenamefont {Gerardot}}]{Kuklewicz2013PRB}%
  \BibitemOpen
  \bibfield  {author} {\bibinfo {author} {\bibfnamefont {L.}~\bibnamefont
  {Sapienza}}, \bibinfo {author} {\bibfnamefont {R.~N.~E.}\ \bibnamefont
  {Malein}}, \bibinfo {author} {\bibfnamefont {C.~E.}\ \bibnamefont
  {Kuklewicz}}, \bibinfo {author} {\bibfnamefont {P.~E.}\ \bibnamefont
  {Kremer}}, \bibinfo {author} {\bibfnamefont {K.}~\bibnamefont {Srinivasan}},
  \bibinfo {author} {\bibfnamefont {A.}~\bibnamefont {Griffiths}}, \bibinfo
  {author} {\bibfnamefont {E.}~\bibnamefont {Clarke}}, \bibinfo {author}
  {\bibfnamefont {M.}~\bibnamefont {Gong}}, \bibinfo {author} {\bibfnamefont
  {R.~J.}\ \bibnamefont {Warburton}},\ and\ \bibinfo {author} {\bibfnamefont
  {B.~D.}\ \bibnamefont {Gerardot}},\ }\href
  {https://doi.org/10.1103/PhysRevB.88.155330} {\bibfield  {journal} {\bibinfo
  {journal} {Physical Review B}\ }\textbf {\bibinfo {volume} {88}},\ \bibinfo
  {pages} {155330} (\bibinfo {year} {2013})}\BibitemShut {NoStop}%
\end{thebibliography}
%

\end{document}